\documentclass[12pt]{article}
\usepackage[cp1251]{inputenc}
\usepackage{amsfonts}
\usepackage[english]{babel}
\usepackage{eufrak}
\usepackage{setstack}
\usepackage{cite}
\usepackage{amssymb}
\title{The Significance of Non-ergodicity Property of Statistical Mechanics Systems for Understanding Resting State of a Living Cell}
\begin{document}
\author{D.V. Prokhorenko \footnote{Institute of Spectroscopy, Russian Academy of Sciences, 142190 Moskow Region, Troitsk,  prokhordv@yandex.ru}, V.V. Matveev
\footnote{Institute of Cytology, Russian Academy of Sciences, Saint
Petersburg, vladimir.matveev@gmail.com}}

 \maketitle
\begin{abstract}
\maketitle A better grasp of the physical foundations of life is
necessary before we can understand the processes occurring inside a
living cell. In his physical theory of the cell, American
physiologist Gilbert Ling introduced an important notion of the
resting state of the cell. He describes this state as an independent
stable thermodynamic state of a living substance in which it has
stored all the energy it needs to perform all kinds of biological
work. This state is characterised by lower entropy of the system
than in an active state. The main contribution to this reduction in
entropy is made by the cellular water (the dominant component with a
concentration of 14 M) which remains in a bound quasi-crystallised
state in a resting cell. When the cell becomes active the water gets
desorbed and the system's entropy goes up sharply while the  free
energy of the system decreases as it is used up for biological work.
However, Ling's approach is primarily qualitative in terms of
thermodynamics and it needs to be characterised more specifically.
To this end, we propose a new thermodynamic approach to studying
Ling's model of the living cell (Ling's cell), the centrepiece of
which is the non-ergodicity property which has recently been proved
for a wide range of systems in statistical mechanics [7]. In many
ways this new thermodynamics overlaps with the standard
quasi-stationary thermodynamics and is therefore compatible with the
principles of the Ling cell, however a number of new specific
results take into account the existence of several non-trivial
motion integrals communicating with each other, whose existence
follows from the non-ergodicity of the system (Ling's cell). These
results allowed us to develop general thermodynamic approaches to
explaining some of the well-known physiological phenomena, which can
be used for further physical analysis of these phenomena using
specific physical models.
\end{abstract}

\newpage

 \section{Introduction}
The living state of a substance has always attracted the attention
of physicists. And indeed, only a clear understanding of the
thermodynamic characteristics of the living matter can give us
insight into the processes that occur in living organisms. However,
despite the fact that there is a lot of interest in this problem,
this field can't be said to be developing by leaps and bounds.

Before proceeding with our analysis we have to establish its
boundaries and conditions.  There are two very different approaches
to the thermodynamics of living systems: one based on thermodynamics
of equilibrium process [1,2,3,4] and the other on thermodynamics of
non-equilibrium processes [5]. Shroedinger's research [6] differs
from both of these.

An obscure Russian scientist of Hungarian descent Ervin Bauer [4]
was probably the first to suggest that the living state should be
regarded as an unstable equilibrium. Treating the physical state of
living substance in this way allowed him to draw a number of
interesting conclusions and generalizations, but on the whole the
most part of research was purely theoretical.

According to Ling [1,2,3], whose position is of special interest to
us, the minimal cell in the physical sense is a complex comprising
protein in an unfolded configuration and water with ions , water and
ions. The most significant characteristic of this complex is the
state of water in it; it is absorbed by the protein in the form of a
multi-layered structure that surrounds it along the entire length of
the polypeptide. This 'coat' consisting of water molecules is
stabilized by hydrogen links that are stronger than the hydrogen
links in volumetric water. This increase in strength is a result of
an increase in the dipole moment of the water molecules under the
influence of other dipoles that are stronger than water such as the
functional groups in the peptide link bound (NH and CO). The
polarization of water molecules explains both their strong binding
by the polypeptide frame of the protein and the multi-layered
absorption of water on the surface of the unfolded protein.
According to Ling, almost all the water in a the cell is in a bound
state.  Because, in terms of the number of molecules, water is the
most abundant compound inside the cell, its transition into a
quasi-crystal state results in a significant fall in the entropy of
the cell. This lower entropy is what brings about the rise in the
amount of free energy of the resting living cell.

According to Ling [1,2,3], whose position is of special interest to
us, the minimal cell in the physical sense is a complex comprising
protein in an unfolded configuration, water and ions. The most
significant characteristic of this complex is the state of water in
it; it is absorbed by the protein in the form of a multi-layered
structure that surrounds it along the entire length of the
polypeptide. This 'coat' consisting of water molecules is stabilized
by hydrogen links that are stronger than the hydrogen links in
volumetric water. This increase in strength is a result of an
increase in the dipole moment of the water molecules under the
influence of other dipoles that are stronger than water such as the
functional groups in the peptide bound (NH and CO). The polarization
of water molecules explains both their strong binding by the
polypeptide frame of the protein and the multi-layered absorption of
water on the surface of the unfolded protein.  According to Ling,
almost all water in the cell is in a bound state.  Because, in terms
of the number of molecules, water is the most abundant compound
inside the cell, its transition into a quasi-crystal state results
in a significant fall in the entropy of the cell. This lower entropy
is what brings about the rise in the amount of free energy of the
resting living cell.

The introduction of the concept of a resting state is one of Ling's
achievements. It's this state that is used as the reference point
for all the physical and chemical processes that take place inside
the cell. When a cell is activated by an external stimulant or some
other signal, it changes its state from resting to active. The
active state is characterized by the disintegration of the
water-protein-ions complex.  The bound water breaks free and the
system's entropy increases. The free energy of the resting state is
released and is used up for all kinds of biological work. This is
the Ling model of the living cell (Ling's cell) that will be the
focus of our analysis.

According to Ling, a cell can remain resting without exchanging
energy or substances with the external environment. The cell just
maintains diffusion equilibrium with the. This view directly
contradicts Prigozhin's approach according to which a living cell
can only maintain its organization as long as it keeps exchanging
substance and energy with the environment on a continuous basis. In
other words, a cell can be likened to a burning candle flame; the
flame will remain 'alive' only as long as there is sufficient supply
of fuel and oxidizer.

Ling believes that such understanding of the thermodynamics of life
is completely inadequate in the case of a living cell. His
calculations demonstrate that if a continuous inflow of energy was
really necessary for the experimentally observed exchange of Na+
ions between a resting cell and the environment (as is postulated in
the traditional mechanism), the cell would simply be incapable to
produce the necessary amounts of energy [3] and therefore the
universally adopted model of ion transport contradicts the energy
preservation law. Ling's other argument proceeds as follows; if a
living cell and a burning candle were to be frozen to the
temperature of liquid nitrogen, both the flame and the life in the
cell will 'go out', but if they're heated back to room temperature,
the flame won't start burning again, but the life processes in the
cell will resume.

The contradictions between the thermodynamic approaches to the
phenomenon of life are so pronounced that the need for further
research in this field is self-evident. The purpose of this paper is
the demonstrate that the property of non-ergodicity that has
recently been proved for a large number of systems in statistical
mechanics can help better understand the nature of the resting state
of a Ling's cell and supports his understanding of the living cell's
thermodynamics.

One feature of this approach is that it suggests that the resting
state of a Ling's cell should be considered to be a non-equilibrium
stationary state, whose existence is a direct consequence of the
non-ergodicity property that we postulate for Ling's cells. In this
approach thermodynamics of non-equilibrium stationary states must be
constructed (analogous to the standard quasi-stationary
thermodynamics) to explain why biological work becomes possible in
the context of the proposed approach (biological work here means any
changes in the cell that have a biological significance and that use
up energy, for instance muscle contraction). There is no real
contradiction between Ling's stationary resting state and the
obviously continuous metabolism necessary to maintain life, because
a real cell constantly changes its state from active to resting and
back. Metabolism and energy are needed to go back to a resting state
rather than maintain it.

Non-ergodicity means that there exist non-trivial first integrals of
the system (i.e. there are values that are invariant under the
Heisenberg and Hamilton motion equations). Because by definition we
consider these first integrals to be real, then, in the case of
quantum mechanics, they must be represented by self-adjoint
operators and be experimentally observable values. It's then only
natural to ask how come they can only be observed in biological
systems (as is shown below) as well as in liquid helium and
superconductors, systems that are as far from biological as can be?
When answering this question we come across a certain mathematical
similarity between the super-fluid state of helium and the resting
state of the Ling cell, and this similarity helps us better
understand the physics of the living state.

The main result of this paper is that by looking at the Ling cell as
a non-ergodic system we were able to propose a common physical
mechanism for various physiological phenomena, which were previously
explained with the help of separate mechanisms barely related to
each other. The main goal of physics in physiology must be to find
out the thermodynamic nature of an active living cell, i.e. the
source of all the manifestations of life. With this goal in mind we
only discuss some of the characteristics of the Ling cell and
provide only a most general physical description for them. We
demonstrate, for example, that when a Ling's cell is activated it
emits heat rather than absorbs it. When we consider the properties
of the physical model we use, it becomes perfectly clear why it is
that unfolded proteins that make up the structural foundation of a
resting Ling's cell, begin to fold when it goes active, why
potassium ions exit the cell into the environment and why an active
cell changes its size (usually it shrinks) and what makes a cell
dead. It is not our goal in this paper to compare the results we
obtain for the Ling's cell with the properties of a real living
cell.

\section{Non-ergodicity of statistical mechanics systems.}

In this section we first give the definition of ergodicity in
statistical mechanics. We then formulate the main result of [7] and
its classical analogue. After that we demonstrate how non-ergodicity
follows from this result on the classical level.

\textbf{Definition}. Suppose that a quantum system is described by
Hamiltonian \(H\) and some set of self-adjoint integrals of motion
\(K_1\),...,\(K_l\) commuting to each other. We say that the system
is ergodic with respect to the integrals \(K_i\),...,\(K_l\), if
each dynamical variable commuting with  \(H\), \(K_1\),...,\(K_l\),
is their function, or in other words the joint spectrum of operators
\(H\), \(K_1\),..., \(K_l\)  is simple.

The classical version of this definition can be easily obtained by
replacing the word commutator with a Poisson bracket.

In our case of Bose gas, apart from the Hamiltonian, there are the
following trivial commuting first integrals: impulse \(\vec{P}\) and
the number of particles \(N\). We will be looking at ergodicity
relative to this set of first integrals.

Let us now formulate the main result of our work [7]. Let
\(\Psi(x)\) and \(\Psi^+(x)\) be secondary quantized wave functions
of Bose gas that satisfy the canonical commutative relations

\begin{eqnarray}
{[\Psi(x),\Psi(x')]=[\Psi^+(x),\Psi^+(x')]=0},\nonumber\\
{[}\Psi(x),\Psi^+(x'){]}=\delta(x-x'),
\end{eqnarray}
where the brackets designate a commutator and \(\delta(x-x')\)-
means Dirak \(\delta\) --- function of a vector argument. An algebra
generated by secondary-quantized wave functions is called the
algebra of canonical commutative relations. The main result of [7]
can be formulated as the following theorem.

\textbf{Theorem 1.} There exist a linear functional
\(\langle\cdot\rangle\)  on the algebra of canonical commutative
relations (within a formal perturbation theory applied to an
interaction constant) such that:

a) It is stationary, i.e. \(\langle[H,v]\rangle=0\), where \(H\) is
a Hamiltonian of the system and  \(v\)  is an arbitrary element of
the canonical commutative relations algebra.

b)\(\langle\cdot\rangle\)  is translation-invariant.

c))\(\langle\cdot\rangle\) Commutates with the operator of the
number of particles \(N\), i.e. \({([N, v]}) = 0\), where \(N\) is
the operator of the number of particles in the system and \(v\) is
an arbitrary element of the algebra of commutative relations.

d)  Satisfy to the week cluster property, i.e.

\begin{eqnarray}
\lim_{|{a}|\rightarrow\infty}\int \langle\Psi^\pm(t,{x}_1+\delta_1
e_1{a})
...\Psi^\pm(t,{x}_n+\delta_n e_1 {a} )\rangle f({x}_1,...,{x}_n)d^3x_1...d^3x_n\nonumber\\
=\int\langle\Psi^\pm(t,{x}_{i_1})...\Psi^\pm(t,{x}_{i_k})\rangle
\langle\Psi^\pm(t,{x}_{i_k})...\Psi^\pm(t,{x}_{i_n})\rangle
\nonumber\\
\times f({x}_1,...,{x}_n)d^3x_1...d^3x_n,
\end{eqnarray}
where \(\delta_i\in\{1,0\},\;i=1,2...n\) and
\begin{eqnarray}
i_1<i_2<...<i_k,\nonumber\\
i_{k+1}<i_{k+2}<...<i_n,\nonumber\\
\{i_1,i_2,...,i_k\}=\{i=1,2...n|\delta_i=0\}\neq\emptyset,\nonumber\\
\{i_{k+1},i_{k+2},...,i_n\}=\{i=1,2...n|\delta_i=1\}\neq\emptyset.
\end{eqnarray}

\(f(x_1,...,x_n)\) - is a test function (i.e. it is sufficiently
smooth and it decreases sufficiently fast at infinity with all its
derivatives), \(e_1\) is a unit vector parallel to the x axis.

e)\(\langle\cdot\rangle\) is not describe the Gibbs distribution.

Before proving the non-ergodicity of the system from this result, we
will formulate its classical analogue. In classical statistical
mechanics a system containing \(N\) particles \(N\rightarrow
+\infty\), is described by a distribution function of the form
\(\rho((x_1,p_1),...,(x_N,p_N))\), where \((x_i,p_i)\) are the
coordinates and impulse of the \(i\)-th particle \(i = 1,...,N\).
The distribution function  \(\rho\) is assumed to be symmetrical
relative to the permutation of its arguments \((x_1,p_1),...,
(x_N,p_N)\).

Also partial distribution functions are introduced:
\begin{eqnarray}
\rho_0=1,\nonumber\\
\rho_1(x_1)=V \int dx_2dp_2...dx_Ndp_N
\rho((x_1,p_1)...(x_N,p_N)),\nonumber\\
\rho_2((x_1,p_1),(x_2,p_2))=V^2 \int dx_3dp_3...dx_Ndp_N
\rho((x_1,p_1)...(x_N,p_N)),\nonumber\\
......................................................................................................
\label{DISTR}
\end{eqnarray}
\(V\) is a volume of the system. Partial distribution functions
 \(\rho_0=1,\rho_1,...,\rho_n...\) satisfy the following consistency
 condition:
\begin{eqnarray}
\rho_n((x_1,p_1),...,(x_n,p_n))=\nonumber\\
=V^{-m} \int
dx_{m+1}dp_{m+1}...dx_{n+m}dp_{n+m}\rho_n((x_1,p_1),...,(x_{n+m},p_{n+m})).
\end{eqnarray}
If the number of particles in the system is variable, then in a
similar way to (\ref{DISTR}), we can also define partial
distribution fucntions. They will also satisfy the consistency
conditions if \(N\) approaches infinity. If we take the main result
of [7] and pass it through the limit we get the following theorem.

\textbf{Theorem 1'}. There exists such  a state \(\rho\) (not
necessarily positively defined and possibly with a varied number of
particles) of a gas consisting of particles with weak interaction,
that

a) This state is stationary, i.e. it doesn't change with time due to
its motion equations.

b)  This state is translation-invariant.

c) This state satisfies the weak cluster property, i.e.  \(\forall
n=1,...,\infty\) and  \(\forall k=1,...,n\)

\begin{eqnarray}
\lim
\limits_{|l|\rightarrow\infty}\rho_n((x_1,p_1),...,(x_k,p_k),(x_{k+1}+le_1,p_{k+1}),...,(x_{n}+le_1,p_{n}))=\nonumber\\
=\rho_k((x_1,p_1),...,(x_k,p_k))\rho_{n-k}((x_{k+1},p_{k+1}),...,(x_{n},p_{n}))
\end{eqnarray}

d)  The state \(\rho\) cannot be described with a Gibbs
distribution.

Let us finally demonstrate how this result leads to the
non-ergodicity of our system. To simply things we'll only be
considering the case of classical mechanics. For a detailed
discussion of the quantum version of this result see [7].

Let us suppose that our system is ergodic, i.e. that its energy and
impulse form a maximal set of first integrals in involution. Two
dynamic variables are said to be in involution if their Poisson
bracket equals zero. From here on for the sake of brevity, we'll be
simply talking about energy rather than energy, impulse and the
number of particles. Let  \(f\) be a distribution function for our
system in a phase space that corresponds to the state we talked
about before. Then because of the system's assumed ergodicity it
must be a function of only energy \(f=f(E_\Gamma)\) (\(\Gamma\) - is
a point in the phase space) and can be presented as a superposition
of micro-canonical distributions:

\begin{eqnarray}
f(E_\Gamma)=\sum \limits_\alpha c_\alpha \delta(E_\Gamma-E_\alpha),
\end{eqnarray}

The summation in this context is understood in the broadest possible
sense and can be infinite (in other words it can be replaced with
integration).

Let 1 be a sufficiently large but finite subsystem of our system and
let 2 -   be a subsystem obtained from 1 by shifting it along the
\(x\) axis by a distance of \(l\). Let 12 -   be a union of systems
1 and 2. Let \(\Gamma_1\), \(\Gamma_2\) and \(\Gamma_{12}\)
designate points in the phase space for our systems 1, 2 and 12
respectively. Let \(f_1(\Gamma_1)\), \(f_2(\Gamma_2)\) and
\(f_{12}(\Gamma_{12})\) be distribution functions for systems 1, 2
and 12 respectively. Then using the same method that is used for
obtaining a canonical distribution from a microcanonical one, for
\(|l|=\infty\)  we can find that:

\begin{eqnarray}
 f_{12}=\sum c_\alpha
d_\alpha\frac{e^{-\frac{E_{\Gamma_1}}{T_\alpha}}}{Z_\alpha}\frac{e^{-\frac{E_{\Gamma_2}}{T_\alpha}}}{Z_\alpha}.\label{I}
\end{eqnarray}

Here \(d_\alpha>0\) - is a weight multiplier, \(E_{\Gamma_1}\) and
\(E_{\Gamma_2}\) - are the energies of subsystems 1 and 2,
\(T_\alpha\) - is the temperature corresponding to energy
\(E_\alpha\), and \(Z_\alpha\) - is the statistical sum:

\begin{eqnarray}
Z_\alpha=\int d \Gamma_1 e^{-\frac{E_{\Gamma_1}}{T_\alpha}}.
\end{eqnarray}
But from the weak cluster property it follows that:
\begin{eqnarray}
f_{12}=f_1f_2.
\end{eqnarray}
Therefore it is follows from (\ref{I}) that all \(c_\alpha=0\)
except one and:
\begin{eqnarray}
f(E)=c\delta(E-E_0),
\end{eqnarray}
(for some \(c\) and \(E_0\)), i.e. the whole system is described by
a microcanonical distribution while any of its sufficiently large
but finite subsystems is described by a canonical Gibbs
distribution, which contradicts the fact that our state is not a
Gibbs state. This contradiction proves that our system is
non-ergodic.

\section{Discussion of Non-ergodicity} Because not enough research has
bee conducted into the specific features of the dynamics of
statistical mechanics systems connected with non-ergodic property,
the discussion in this section will be informal and hypothetical in
places.

\textbf{Boltzmann's hypothesis.} Boltzmann hypothesis (1871)
postulates that in statistical mechanical systems the dynamic
variables that are average in time are equal to the dynamic
variables that are average across the (micro) canonical ensemble. We
will only consider the quantum mechanics case. The discussion
presented here was taken from [7]. Before we can formulate the
Boltzmann hypothesis for a Bose gas with weak pair interaction, some
designations need to be introduced. Let  \(\langle
a\rangle_{\beta,\vec{v},\mu}\) be the Gibbs state corresponding to
the inverse temperature \(\beta\), the system velocity \(\vec{v}\)
and its chemical potential  \(\mu\), i.e.
\begin{eqnarray}
 \langle a\rangle_{\beta,\vec{v},\mu}=\frac{1}{Z_{\beta,\vec{v},\mu}} {\rm tr \mit} (a
 e^{-\beta(H-\mu N+\vec{v}\vec{P})}),
\end{eqnarray}
where \(a\) is an arbitrary element from the canonical commutative
relations algebra, \(H\) is the Hamiltonian of the system, \(N\) is
the operator of the number of particles in the system, \(\vec{P}\)
is the operator of the system impulse and \(Z\) is the so called
large statistical sum, i.e.
\begin{eqnarray}
 Z_{\beta,\vec{v},\mu}={\rm tr \mit} (e^{-\beta(H-\mu N+\vec{v}\vec{P})}).
 \end{eqnarray}

Let  \(V'_G\) be a space of all linear functionals on the algebra of
canonical commutative relations, each of which is a superposition of
Gibbs states.

Let  \(\langle\cdot\rangle\)  be is a translation-invariant
functional on the algebra of canonical commutative relations such
that  \(\forall t \in \mathbb{R}\) \(\langle e^{itH}(\cdot)
e^{-itH}\rangle\) is a well defined linear functional. The Boltzmann
hypothesis states that in this case there will be a linear
functional \(\langle\cdot\rangle' \in V'_G\) such that for any
element \(a\) from the algebra of commutative relations:
\begin{eqnarray}
\lim \limits_{T\rightarrow +\infty}\frac{1}{T} \int
\limits_{0}^{T}\langle e^{itH} a e^{-itH}\rangle dt=\langle
a\rangle'. \label{Bol}
\end{eqnarray}

Let us demonstrate that if formulated in this way the Boltzmann
hypothesis is wrong. Theorem 1 states that there exists a
translation invariant stationary linear functional on the algebra of
canonical commutative relations and that said functional is not a
superposition of Gibbs functions. However, this functional is built
only in the form of a formal power series on the interaction
constant and nothing is known about whether this series has a finite
sum or not. [7], however, demonstrates that there is a way around
the issue of whether this series converges or not and that there
exists a stationary translation invariant functional on the algebra
of canonical commutative relations \(\langle\cdot\rangle_s\)  that
does not belong to \(V'_G\). If this composite function is
substituted for  \(\langle\cdot\rangle\) in the left side of
equation  (\ref{Bol}), then the left side itself will equal
\(\langle\cdot\rangle_s\) and from equation (\ref{Bol}) it will then
follow that \(\langle\cdot\rangle_s\) is superposition of Gibbs
states. This contradiction shows that the Boltzmann hypothesis does
not hold.

\textbf{Discussion of the relation between a system's behaviour on
its boundaries and its tendency towards thermodynamic equilibrium.}
Non-ergodicity means that Bose gas (with weak pair interaction) will
not tend towards thermodynamic equilibrium in an infinite volume.
Consequently to prove the tendency towards thermodynamic equilibrium
we have to take into account its behaviour on the boundaries. In
fact if the role played by the boundaries is disregarded we may end
up with an infinite system.

In [7] the role of boundaries is illustrated with Bogolubov's
derivation of Boltzmann's kinetic equation. Bogolubov's programme
for deriving equations considers an infinite strengthening chain of
equations for partial distribution functions (BBGKI chain)
equivalent to Luiville's equations, and then closed kinetic
equations are obtained by selecting the suitable conditions for
breaking the chain (correlation breaking conditions ). The first
member of this chain has the form:
\begin{eqnarray}
\frac{\partial}{\partial
t}\rho_1((x_1p_1)|t)+\frac{{p}}{m}{\nabla}\rho_1((x_1,p_1)|t) +\int
dx_2dp_2 \frac{{p_2}}{m}\frac{\partial}{\partial
{x}_2}\rho_2((x_1,p_1),(x_2,p_2)|t)=\nonumber\\
=\int dx_2dp_2\frac{\partial V(x_1-x_2)}{\partial
{x}_1}\frac{\partial \rho_2((x_1,p_1),(x_2,p_2)|t)}{\partial p_1}.
\label{II}
\end{eqnarray}
Here \(t\) is time, \(m\) is the mass of the particles, and \(V\) is
the potential of pair interaction. An integral over the coordinates
in the last term in the equation on the left (\ref{II}) may be
transformed into an integral along the boundary of a
three-dimensional space using the Gauss theorem, and usually this
fact is used as the basis for assuming that this integral equals to
zero. However in [7] it is demonstrated if the last term on the left
side of  (\ref{II}) is not discarded and Bogolubov's strategy for
deriving a Boltzmann equation is followed (if translation-invariant
distributions are considered from the start), then this term will
eventually cancel out with the collision integral and the system
will have no kinetic evolution. Even though, it is possible that
this effect is purely mathematical, it does make one wonder about
the role that the system's behaviour on its boundaries plays in
proving its tendency towards thermodynamic equilibrium.

Although it is now clear that the system's behaviour on the boundary
must be taken into account when proving its tendency towards
thermodynamic equilibrium,  the following question now arises: if
the linear sizes of the system equal   \(\sim L\) and tend to
infinity, then the volume of the system will b
ehave like \(L^3\)
while the surface area of its boundary will equal \(L^2\)  which
makes one think that the influence of the boundaries must be very
insignificant.

A similar situation can be observed in the theory of second order
phase transitions, for instance in the theory of ferromagnetism. If,
for instance, you take a two dimensional Izing model at a
temperature below the critical level, and assume that all the spins
are directed upwards along the boundaries of the crystal, then
Peierls estimates (see [8]) show that in the thermodynamic limit
there will be a non-zero magnetization. Second order phase
transitions are often related to breaking of some symmetry. In the
case of the Izing model, the symmetry that is broken is the symmetry
relative to the change in the direction of all the spins in the
grid. In the case of kinetic evolution there is also as symmetry
that gets broken, it's the symmetry relative to the time sign
(Boltzmann's H-theorem). We will have a more detailed discussion of
this similarity later on.

The necessity to take into account a system's behaviour on its
boundaries when proving the system's tendency towards thermodynamic
equilibrium raises an interesting philosophical question. If a
system's behaviour on its boundaries must be taken into account when
proving the system's tendency towards thermodynamic equilibrium,
then it means we have to know about the behaviour of the particles
outside the system's boundaries because those influence the
behaviour of the particles immediately inside the system's
boundaries, in other words we have to know how the environment that
the system is a part of behaves. This in turn means that we have to
know the behaviour of the larger environment around the immediate
environment of the system and so on and so forth. So how do we
escape from this infinite regress?

To answer this question we will consider the following example. It
is a known fact that there are black holes in our universe. Black
holes are known to emit energy but for this phenomenon to be
explained certain boundary conditions have to be assumed, which
essentially are based on the fact that a black hole is formed as a
result of the collapse of a mass body. According to one theorem of
Penrose, this means that at one time in the past there existed a
certain feature of the space time. The surface of a black hole, the
so called event horizon, is the boundary at which the dependence of
the particles inside the system on particles outside the system is
broken. In fact due to the structure of light cones, information on
the event horizon can only fall into the black hole, while the
Hawking radiation is in equilibrium and therefore carries only the
bare minimum of information.

On the other hand one of Penrose-Hawking's theorems [9] (whose
conditions are confirmed by reliable astronomical observations)
postulates that at some time in the past there was a singularity of
space-time which can be equated to the Big Bang. At the same time
Boltzmann also posited (on the basis of his \(H\)-theorem) that
there had to have been a moment in time which can be equated to the
beginning of the universe. In the theorems of Penrose and Hawking
there appears entropy (equal to the surface area of the event
horizon), which increase and is related to the temperature of the
Hawking radiation through a known thermodynamic relation. It follows
from all of this that there is a much deeper link between the
'proof' of Boltzmann's Big Bang and Penrose --- Hawking's
singularity than a simple mathematical similarity.

It is also worth noting that due to the arguments presented above,
the non-ergodicity of Bose gas with weak pair interaction can be
interpreted as the impossibility of deriving macroscopic dynamics
from microscopic dynamics.

The non-ergodicity theorem, as follows from its proof (see [7]), is
applicable to a broad range of systems in statistical mechanics and
we make the assumption that it can also be applied to the Ling's
cell. Therefore, the general results obtained for all non-ergodic
systems can also be applied in this case.

A more detailed study of how the properties of a system's surface
influence the process whereby it achieves thermodynamic equilibrium
can be helpful in improving our understanding of living cells and
the role of the cell's surface in the physiological processes inside
it.

\textbf{Analogy with second order phase transitions.} Let's discuss
in more detailed the aforementioned analogy between the broking of
the symmetry relative to the time sign and second order phase
transitions. We'll do this by discussing the Green-Cubo formula that
describes the linear response of a system to small perturbations.
Let there be \(n\) fluctuating variables in the system
\(x_1,...,x_n\) and let \(\hat{x}_1,...,\hat{x}_n\) be corresponding
quantum mechanics operators. Let us assume that the fluctuations are
sufficiently small and that the system's entropy can be expressed
with the formula
\begin{eqnarray}
S=S_0- \beta_{ij}x_ix_j,
\end{eqnarray}
where \(\beta_{ij}\)  is some positively defined matrix (here and
through the end of this section indexes repeated twice mean
summation). Let \(X_1,...,X_n\) be thermodynamically conjugate
values to  \(x_1,...,x_n\)  i.e.
\begin{eqnarray}
X_k=-\frac{\partial S}{\partial x_k}.
\end{eqnarray}
The equation describing the relaxation of this system look like
this:
\begin{eqnarray}
\dot{x}_i=-\gamma_{ik}X_k,\label{Lars}
\end{eqnarray}

where \(\gamma_{ik}\) - are the so called kinetic coefficients. This
coefficients satisfy Onsanger's symmetry principle:
\begin{eqnarray}
\gamma_{ik}=\gamma_{ki}.
\end{eqnarray}

If the system is affected by external forces \(f_k\;k=1,...,n\), in
other words if the system's Hamiltonian looks like this
\begin{eqnarray}
H=H_0+f_k\hat{x}_k,
\end{eqnarray}
(\(H_0\) is the Hamiltonian of an unperturbed system), then we can
determine the generalized sensitivity of the system
\(\alpha_{ik}(\tau)\) in such a way that
\begin{eqnarray}
x_k(t)=\int \limits_{-\infty}^{+\infty}
\alpha_{ki}(t-\tau)f_i(\tau)d\tau.
\end{eqnarray}

Let's use \(\tilde{\alpha}_{ik}(\omega)\) to designate the Fourier
transform of the generalized sensitivity
\begin{eqnarray}
\tilde{\alpha}_{ik}(\omega)=\int e^{i\omega t} \alpha_{ik}(t) dt.
\end{eqnarray}
The generalized sensitivity can be found from the Green --- Cubo
formula:
\begin{eqnarray}
\alpha_{ik}(t)=-i\theta(t)\langle[\hat{x}_i(t),\hat{x}_k(0)]\rangle,
\label{Vos1}
\end{eqnarray}
where \(\theta(t)\) is a Heaviside step function that equals 1 when
its argument is positive and zero if its negative and
\(\langle\cdot\rangle\) is the averaging over the equilibrium state
of an unperturbed system. It has to be noted, however, that when the
Green---  Cubo formula is applied the system is assumed to be in
equilibrium with   \(t=-\infty\). If we assume, on the contrary,
that the system is in equilibrium with \(t=\infty\), then by
transforming the time in the Green Cubo --- formula, we will get the
following formula for the generalised sensitivity
\begin{eqnarray}
\alpha'_{ik}(t)=i\theta(-t)\langle[\hat{x}_i(t),\hat{x}_k(0)]\rangle.
\end{eqnarray}
It can be easily seen that
\begin{eqnarray}
\tilde{\alpha}'_{ik}(\omega)=(\tilde{\alpha}_{ik}(\omega))^\star,
\end{eqnarray}
where the asterisk means complex conjugation. If, however, we start
with an approximation in which relaxation equations (\ref{Lars})
hold, then it can be shown that there exists the following relations
between the matrices \(\alpha,\beta\) and \(\gamma\):
\begin{eqnarray}
\tilde{\alpha}_{ik}(\omega)=\frac{1}{T}(\beta_{ik}-i\omega\gamma_{ik}^{-1})^{-1}.\label{Vos2}
\end{eqnarray}
It can be seen from this formula that if \(t\) is replaced with
\(-t\) and vice versa in the Green -- Cubo formula, then
\(\gamma_{ik}\)  will become \(-\gamma_{ik}\)  and will no longer
describe kinetic evolution.

So, how can we choose between the requirement for equilibrium with
\(t=-\infty\) and with \(t=+\infty\), in other words how can we
determine whether the equation with a retarded commutator
(\ref{Vos1}) or the equation with a advanced commutator (\ref{Vos2})
should be used for the generalized sensitivity of the system?

It should be noted that a similar problem arises in classical
electrodynamics when working out the formula for retarded
potentials. A vector potential \(A^\mu,\;\mu=0,1,...,3\) in
classical electrodynamics satisfies the heterogeneous wave equation:
\begin{eqnarray}
\Box A^\mu=J^\mu,
\end{eqnarray}
where
\begin{eqnarray}
\Box=\frac{\partial^2}{(\partial x^0)^2}-\frac{\partial^2}{(\partial
x^1)^2}-\frac{\partial^2}{(\partial
x^2)^2}-\frac{\partial^2}{(\partial x^2)^3},\label{Vawe}
\end{eqnarray}
and  \(J^\mu\) is a four-vector of current, satisfying the
continuity equation:
\begin{eqnarray}
\partial_\mu J^\mu=\frac{\partial}{\partial x^0}J^0+\frac{\partial}{\partial
x^1}J^1+\frac{\partial}{\partial x^2}J^2+\frac{\partial}{\partial
x^3}J^3=0.
\end{eqnarray}
Where \(x_0\) is time, and \(x_1,x_2,x_3\) are the spacial Euclidean
coordinates.

The solution to equation (\ref{Vawe}) usually has the form:
\begin{eqnarray}
A^\mu(x)=\int D(x-y)J^\mu(x)d^4x, \label{Zap}
\end{eqnarray}
where \(D(x)\) is the fundamental solution to the wave equation:
\begin{eqnarray}
\Box D(x)=\delta(x),
\end{eqnarray}
with a support in the upper light cone:
\begin{eqnarray}
{\rm supp \mit} D(x)\subset \bar{V}^+= \{x|x_0\geq 0,\;x^2\geq 0\}.
\end{eqnarray}
There exists a single fundamental solution that satisfies this
property. Formula (\ref{Zap}) is called the formula of retarded
potentials. The similarity to the Green---Cubo formula becomes even
closer if we remember that the vacuum mean of the commutators of the
secondary quantum vector potential in electrodynamics is presented
in the following form
\begin{eqnarray}
\langle
0|[A^\mu(x),A^\nu(y)]|0\rangle\theta(x^0-y^0)=ig^{\mu\nu}D(x-y),
\end{eqnarray}
where  \(g^{\mu\nu}\) is a metric tensor:
\begin{eqnarray}
g^{\mu\nu}=0,\;{\rm if \mit}\; \mu\neq\nu,\nonumber\\
g^{00}=1,\;g^{ii}=-1,\;i=1,2,3.
\end{eqnarray}
If in (\ref{Vawe}) we switch to the Fourier transform in time, then
we will get:
\begin{eqnarray}
\{\omega^2+\frac{\partial^2}{(\partial
x^1)^2}+\frac{\partial^2}{(\partial
x^2)^2}+\frac{\partial^2}{(\partial
x^2)^3}\}A^\mu(\omega,\vec{x})=-J^\mu(\omega,\vec{x}).
\end{eqnarray}
The solution to this equation, corresponding to the formula of
delayed potentials, will have the form:
\begin{eqnarray}
A^\mu(\omega,\vec{x})=-\frac{1}{4\pi}\int
\frac{e^{i\omega|x-y|}}{|x-y|}J^\mu(\omega, \vec{y})d^3\vec{y}.
\end{eqnarray}
If \(S_R\)  is a sphere with a radius \(R\) and its center at zero
and \(R\rightarrow \infty\), then  \(A^\mu(\omega,\vec{x})\) on this
sphere will satisfy the Sommerfeld radiation condition:
\begin{eqnarray}
\frac{\partial}{\partial\vec{n}}A^\mu(\omega,\vec{x})-i\omega
A^\mu(\omega,\vec{x})=O(\frac{1}{R^2}),
\end{eqnarray}
where \(\frac{\partial}{\partial\vec{n}}\) means differentiation
along the external normal to sphere \(S_R\). By applying a inverse
Fourier transform to this formula, we will find the condition that
must be met by  \(S_R\), to ensure that a retarded commutator is
selected in the formula for retarded potentials:
\begin{eqnarray}
\frac{\partial}{\partial\vec{n}}A^\mu(x)+\frac{\partial}{\partial
x^0} A^\mu(x)=O(\frac{1}{R^2}). \label{Zom}
\end{eqnarray}
The selection of a advanced commutator would result in the following
formula
\begin{eqnarray}
\frac{\partial}{\partial\vec{n}}A^\mu(x)-\frac{\partial}{\partial
x^0} A^\mu(x)=O(\frac{1}{R^2}).
\end{eqnarray}
Thus we can see that a advanced or a retarded commutator in
Green---Cubo formulas can be selected (which is necessary to specify
the direction of time), at least in electrodynamics, by applying
appropriate boundary conditions on the infinitely removed boundary
of a three dimensional space.  This approach is in full accord with
our ideas about the role that the boundaries of a system play in
proving its tendency towards thermodynamic equilibrium. It should
also be noted that boundary condition (\ref{Zom}) imposes a limit on
the flow of information (this interpretation is also possible):
information can only go out of the system (flow outside). This
restriction on the direction of the flow of information is realized,
for instance, on the event horizon of a black hole due to the
structure of the light cones on it, as has already been noted above.

Let us go back to our analogy with ferromagnetism (Izing model). We
said earlier that non-zero magnetization can be obtained by changing
the direction of the spins on the boundary of the grid. However,
non-zero magnetization can also be achieved in another way. We can
introduce a fictitious infinitesimally small external field, which
is expressed by adding the following expression to the system's
Hamiltonian
\begin{eqnarray}
h\sum \limits_{p} \sigma_p,
\end{eqnarray}
where the summation is done for all the nodes \(p\) of the grid, and
\(\sigma_p\) is the spin in the node of the grid with  \(\sigma_p\in
\{-1,1\}\). Then the system passes to an unlimited volume and after
that \(h\) starts approaching zero, however it always remains
positive or negative. The means calculated in this way are called
quasi-means (Bogolubov's quasi-means). It turns out that non-zero
magnetization occurs in this case as well (and it's equal to the
magnetization that results when the spins on the boundaries are
given a certain direction).

It begs the question whether it is possible to make the right choice
between a advanced and retarded commutator method when working out a
Green---Cubo formula, by analogy with Bogolubov's quasi-means? The
answer to this question is positive and here is why. Let \(H(t)\) be
a Hamiltonian of the system dependent on time:
\begin{eqnarray}
\hat{H}(t)=\hat{H}_0+\hat{V}f(t),
\end{eqnarray}
Let \(f(t)\) be an external and infinitesimally small classical
force changing with time. The von Neumann equation for the density
matrix will look like this :
\begin{eqnarray}
\frac{d}{dt}\rho(t)=-i[\hat{H}(t),\rho(t)]=-i[\hat{H}_0,\rho(t)]-if(t)[\hat{V},\rho(t)].
\end{eqnarray}

If \(f(t)\equiv 0\), \(\rho=\rho_0\) is the density matrix of the
equilibrium of an unperturbed system. Consequently if we're looking
for \(\rho\) in the first order over \(f(t)\), then the following
equation can be used for \(\rho\):
\begin{eqnarray}
\frac{d}{dt}\rho(t)=-i[\hat{H}_0,\rho(t)]-if(t)[\hat{V},\rho_0].
\label{vn}
\end{eqnarray}
The standard method for solving this equation when working out a
Green --- Cubo formula involves choosing a boundary condition
\(\rho(t)\rightarrow \rho_0\) with \(t\rightarrow -\infty\) and
solving  (\ref{vn}) by using the constant variation method. What we
get then is:
\begin{eqnarray}
\rho=\rho_0-i\int \limits_{-\infty}^t
e^{-i\hat{H}_0(t-\tau)}[\hat{V},\rho_0]e^{+i\hat{H}_0(t-\tau)}
f(\tau)d\tau. \label{GK1}
\end{eqnarray}
This expression for the density matrix immediately leads to a
Green---Cubo formula with a retarded commutator. Instead of imposing
initial conditions on the density matrix, we will introduce an
infinitesimally small term \(\varepsilon\rho(t)\),
\(\varepsilon>0\), \(\varepsilon\rightarrow 0\) into the left side
of (\ref{vn}). In other words instead of (\ref{vn}) we will be
solving the following equation for the density matrix:
\begin{eqnarray}
\frac{d}{dt}\rho(t)+\varepsilon\rho+i[\hat{H}_0,\rho(t)]=-if(t)[\hat{V},\rho_0].\label{vn1}
\end{eqnarray}
Applying a Fourier transform to this equation we get:
\begin{eqnarray}
(-i\omega+\varepsilon+i[\hat{H}_0])\tilde{\rho}(\omega)=-i\tilde{f}(\omega)[\hat{V},\rho_0],\label{v1}
\end{eqnarray}
where \([\hat{H}_0]\) designates the following super-operator:
\begin{eqnarray}
[\hat{H}_0]\rho:=[\hat{H}_0,\rho].
\end{eqnarray}
Therefore it follows from (\ref{v1}) that:
\begin{eqnarray}
\rho(\omega)=\frac{-i}{-i\omega+\varepsilon+i[\hat{H}_0]}[\hat{V},\rho_0]f(\omega).
\end{eqnarray}
Since a Fourier transform transforms convolution into product we can
find:
\begin{eqnarray}
\rho(t)=\int
\limits_{-\infty}^{+\infty}\alpha(t-\tau)f(\tau)d\tau,\label{v2}
\end{eqnarray}
where
\begin{eqnarray}
\alpha(t)=\frac{1}{2\pi}\int \limits_{-\infty}^{+\infty} e^{-i\omega
t}\frac{-i}{-i\omega+\varepsilon+i[\hat{H}_0]}[\hat{V},\rho_0]d\omega.\label{v3}
\end{eqnarray}
The super-operator \([\hat{H}_0]\) is the Liouvillian of an
unperturbed system and is therefore self-adjoint (if we define the
scalar product of the operators applicable in the space of states as
 \(\langle\hat{A},\hat{B}\rangle={\rm tr \mit} (\hat{A}^\star\hat{B})\)). Then, calculating
 the integral in (\ref{v3}) through residue we will get:
(\ref{v3})\begin{eqnarray}
\alpha(t)=-i\theta(t)e^{(-i[\hat{H}_0]-\varepsilon)t}[\hat{V},\rho_0]
=\theta(t)e^{-\varepsilon
t}e^{-i\hat{H}_0t}[\hat{V},\rho_0]e^{i\hat{H}_0t}
\end{eqnarray}
Substituting the expression we just found in (\ref{v2}) for
\(\alpha(t)\) and assuming that \(\varepsilon\rightarrow 0\) we will
find:
\begin{eqnarray}
\rho=\rho_0-i\int \limits_{-\infty}^t
e^{-i\hat{H}_0(t-\tau)}[\hat{V},\rho_0]e^{i\hat{H}_0(t-\tau)}
f(\tau)d\tau,
\end{eqnarray}
Thus we've once again got expression (\ref{GK1}) for the density
matrix. Thus we see that boundary conditions for a density matrix
can be chosen using a method similar to Bogolubov's method of quasi
means, i.e. by introducing an appropriate infinitesimally small term
in the equations describing the evolution of the density matrix
through time. This should explain the similarity between spontaneous
breaking of the symmetry in second order phase transitions and the
breaking of the time symmetry in kinetic equations.

\section{Thermodynamics of stationary non-equilibrium states} In
section 1 we said that according to the results obtained in [7],
even the most realistic systems in statistical mechanics are
non-ergodic. It should also be reminded at this stage, that a system
is ergodic if its trivial first integrals (Hamiltonian, impulse,
number of particles) form the maximum set of first integrals in
involution. Hereinafter for the sake of brevity we will use the term
energy to mean the energy and the impulse of a system. From the
non-ergodicity of a system it follows that it must have stationary
states that are not super-positions of Gibbs states. We will call
the states of a system that are super-positions of Gibbs states
equilibrium states.

It therefore makes sense to build thermodynamics of non-equilibrium
stationary states. This thermodynamics will really come in handy in
the next section when we formulate the principle that the substance
inside a living cell is in a non-equilibrium stationary state, which
is the goal of this paper. This section describes how we build this
thermodynamics. The equations we derive in this section will be
applicable to any system in statistical mechanics, including the
Ling's cell. For the sake of simplicity we will only consider the
case of classical mechanics.

Some basic concepts of classical (Hamiltonian) mechanics should be
remembered here. The phase space \(\Gamma\) of a Hamiltonian system
is a \(2n\)-dimensional \(n = 1,2,...\) smooth manifold in which for
any two (good) functions f, g there is a function \((f,g)\)  defined
on their Poisson brackets, which satisfies the following conditions:

1) \((fg, h) = f(g, h) + g(f, h)\) (Leibniz rule)

2) \((f,g) = -(g,f)\) (Anti-symmetry)

3) \((f, (g, h) + (g, (h, f)) + (h, (f,g)) = 0\) (Jacobi identity)

4) For any point  \(x \in \Gamma\), and for any function \(f\):
\({\rm grad \mit} f(x)\neq 0\), there is a function \(g\) on
\(\Gamma\) such that \((f,g)(x)\neq 0\) (non-degeneracy).

\(n\) is called the number of degrees of freedom of the system.

An example of a phase space would be
\(\Gamma=\{(p_1,...,p_n,q_1,...,q_n)\mid p_i,q_i \in \mathbb{R}\}\)
in which the Poisson bracket is defined as:
\begin{eqnarray}
(f,g)=\sum \limits_{i=1}^n(\frac{\partial f}{\partial
p_i}\frac{\partial g}{\partial q_i}-\frac{\partial f}{\partial
q_i}\frac{\partial g}{\partial p_i}). \label{Kan}
\end{eqnarray}
According to the Darboux theorem (see [10]) local coordinates
\(p_1,...,p_n,q_1,...,q_n\) can always be introduces in which the
Poisson bracket will look like (\ref{Kan}). Such coordinates are
called canonical or symplectic.

The evolution of a Hamiltonian system is described by a
one-parameter group of diffeomorphisms (i.e. homeomorphisms which
are smooth with their inverse), \(x\mapsto G_\tau(x),\;\tau \in
\mathbb{R}\) preserving the Poisson bracket (\(\tau\) is time). This
one-parameter group of diffeomorphisms is called a phase flow.

It can be shown that for any phase flow \(G_\tau\) (locally) there
is a function \(H\) on \(\Gamma\), called a Hamiltonian, such that
\begin{eqnarray}
\frac{d}{d\tau}f_\tau=(H,f_\tau),
\end{eqnarray}

Where  \(f_\tau(x):=f(G_\tau(x))\). For more about Hamiltonian
dynamics see ([10]).

Thus, let's assume we have a Hamiltonian system in which in addition
to Hamiltonian \(H\) there is also , \(k = 1,2,3...\) first
integrals \(K_1,...K_k\) in involution, i.e. all the matching
Poisson brackets equal zero: \(\forall i,j=1,...,k\;(K_i,K_j)=0\).
We want to describe a state of our system in which it is in
equilibrium with its environment and in which its Hamiltonian  \(H\)
and the first integrals \(K_1,...,K_k\) take on some specific values
\(E,\;K'_1,...,K'_k\) respectively.

The requirement that the first integrals \(K_1,..., K_k\) must be in
involution can be explained as follows: we want to describe states
in which the integrals of motion \(K_1,...,K_k\) take on specific
values: \(E,\;K'_1,...,K'_k\). But in quantum mechanics (see [11])
simultaneous measurability of the observables means that they are
commutative.  Now it becomes clear why \(K_1,...,K_k\) must be in
involution, if we take into account the fact that the commutator is
the quantum mechanics counterpart of the Poisson bracket (see [12]).

Let us first consider the way the above task is solved with \(k =
0\). The state of a system in statistical mechanics is described by
a distribution function \(\rho(x)\) on the phase space  \(\Gamma\),
i.e.by a function that satisfies the following conditions

a)  \(\forall x \in \Gamma \rho(x)\geq 0\) (Positive definiteness),

b)  \(\int \rho(x)d \Gamma_x=1\) (Normalization).

Here the phase volume is determined in the following way: in
canonical coordinates it looks like
\(d\Gamma=dp_1...dp_ndq_1...dq_n\). It can be demonstrated that this
definition is correct. The observed macroscopic value \(\bar{F}\) of
dynamic variable \(F (x)\) has the form:
\begin{eqnarray}
\bar{F}=\int F(x)\rho(x)d \Gamma_x. \label{SR}
\end{eqnarray}
The role of the distribution function \(\rho(x)\) is assigned to the
so called micro-canonical Gibbs distribution:
\begin{eqnarray}
\rho(x)=c\delta(H(x)-E),
\end{eqnarray}
In which the constant \(c\) is chosen on the basis of the
normalization condition.

It's important to note here that we assumed that the system under
consideration is in thermodynamic equilibrium with its environment.
For this reason \(H(x)\) takes into account interaction with the
particles in the environment and  \(H(x)\) is a function of time and
time is included in \(H(x)\) through the coordinates and impulses of
the particles in the environment. This will be important further
down the road. It has also be remembered, however, that in the
thermodynamic limit these extra terms in \(H(x)\) that account for
the interaction with the environment, can be disregarded.

If the system's Hamiltonian is a function of time, \(H(t)\), then
the system's evolution will be described by a co-cycle, i.e.
\(\forall t_1,t_2\)  there are canonical transformations  (i.e.
transformation that preserves the Poisson bracket)  \(x\mapsto
G_{t_1,t_2}(x)\) meets the co-cycle property:
\begin{eqnarray}
G_{t_1,t_2}G_{t_2,t_3}=G_{t_1,t_3}.
\end{eqnarray}
Any dynamic variable \(f\) is a function of time by definition
\(f_t(x):=f(G_{t,0}(x))\). Now it has to be said that in actuality,
the motion of the system is described by some trajectory through the
phase space and for this reason the left side of (\ref{SR}) has to
be interpreted as a mean over time, or
\begin{eqnarray}
\lim \limits_{T\rightarrow 0}\frac{1}{T} \int
\limits_{0}^{T}F(G_{\tau,0}(x))d\tau=c\int
F(x)\delta(H(x)-E)d\Gamma_x. \label{SR1}
\end{eqnarray}

Equation (\ref{SR1}) does not in any way mean that the system is
ergodic, because from the very beginning we said that our system is
a subsystem of a large system.

A phase flow preserves its phase volume(see [10]). Therefore if  is
the system's distribution function, then \(\rho\) Consequently, over
time the distribution function will evolve into:
\begin{eqnarray}
\rho_t(x)=\rho(G_{0,t}(x)).
\end{eqnarray}
Function  pt meats the conditions of the following differential
equation:
\begin{eqnarray}
\frac{d}{dt}\rho_t=(\rho_t,H(t)).
\end{eqnarray}

Now let us assume that the system's Hamiltonian  \(H_\lambda\) is a
function of some parameter \(\lambda\). We will demonstrate that if
\(\lambda\) undergoes adiabatic changes the micro-canonical
distribution remains the same. Let \(\lambda(t)\) be some function
of time and let \(\varepsilon\) be some small positive number. Let
\(\lambda_\varepsilon(t):=\lambda(\varepsilon t)\). Let \(x\) be a
point in the phase space such that \(H_{\lambda(\varepsilon
t)}(x)=E\). Let \([0,\Delta]\) be a sufficiently long period of time
such that a mean over this period is the same as the mean over the
micro-canonical ensemble. Let us now select such a small
\(\varepsilon\) that will not change very much in this time period.
Let's calculate
\begin{eqnarray}
H_{\lambda_\varepsilon(t)}(G_{t,0}(x))|_{t=\Delta}.
\end{eqnarray}
What we get is:
\begin{eqnarray}
H_{\lambda_\varepsilon(t)}(G_{t,0}(x))|_{t=\Delta}=\nonumber\\
=H_{0}((x))+\int \limits_{0}^\Delta dt
\{\frac{d\lambda}{dt}\frac{\partial
H_{\lambda_\varepsilon(t)}}{\partial\lambda}+(H_{\lambda_\varepsilon(t)},
H_{\lambda_\varepsilon(t)})\}(G_{t,0}(x))=\nonumber\\
=H_{0}((x))+\int \limits_{0}^\Delta
dt\frac{d\lambda}{dt}\frac{\partial
H_{\lambda_\varepsilon(t)}}{\partial\lambda}(G_{t,0}(x))=\nonumber\\
=H_{0}((x))+\Delta \frac{d\lambda}{dt}\overline{\frac{\partial
H_{\lambda_\varepsilon(t)}}{\partial\lambda}}+O(\varepsilon^2)
\end{eqnarray}
Thus with a precision of up to \(\varepsilon^2\) canonical
transformation \(G_{t,0}\) transforms function
\(H_{\lambda_\varepsilon(t)}(x)-E-\Delta
\frac{d\lambda}{dt}\overline{\frac{\partial
H_{\lambda_\varepsilon(t)}}{\partial\lambda}}\) into
\(H_{\lambda_\varepsilon(0)}(x)-E\), i.e. transformation \(G_{t,0}\)
transforms the micro-canonical distribution
\(c\delta(H_{\lambda_\varepsilon(0)}(x)-E)\) into
\(c\delta(H_{\lambda_\varepsilon(t)}(x)-E-\Delta
\frac{d\lambda}{dt}\overline{\frac{\partial
H_{\lambda_\varepsilon(t)}}{\partial\lambda}})\) with a precision of
up to  \(\frac{1}{\varepsilon}\). The function
\(\lambda_\varepsilon(t)\) changes by a finite amount within an
interval of order  \(\frac{1}{\varepsilon}\). Therefore, if we break
down this time period into sub-intervals of duration  \(\Delta\) we
will find that after time period \(\frac{1}{\varepsilon}\) elapses
the form of the micro-canonical distribution will change by a value
of order \(\varepsilon\). If we have an \(\varepsilon\) approaching
zero, we can prove the above contention.

Now we will consider the case when the system has \(k\)- first
integrals in involution. By analogy with the micro-canonical
distribution we will assume that in this case if the values of the
parameters \(E,\;K'_1,...,K'_k\) are fixed the system's distribution
function will have the form:
\begin{eqnarray}
\rho(x)=c\delta(H(x)-E)\prod \limits_{i=1}^k\delta(K_i(x)-K'_i),
\label{MKdis}
\end{eqnarray}
where\(c\) is a normalizing multiplier. First of all let us note
that this definition is correct, i.e. the form of the distribution
will not change if the integrals  \(K_1,..., K_k\) are replaced with
the functions \(\hat{K}_1,...,\hat{K}_k\) of these integrals and the
Hamiltonian. In fact:
\begin{eqnarray}
c\delta(H-E)\prod
\limits_{i=1}^k\delta(\hat{K}_i-\hat{K}'_i)=\nonumber\\
=c\delta(H-E)\prod \limits_{i=1}^k\delta(K_i-K'_i)\frac{D(H,
K_1,...K_k)}{D(H, \hat{K}_1,...\hat{K}_k)},
\end{eqnarray}
where
\begin{eqnarray}
\frac{D(H, K_1,...K_k)}{D(H, \hat{K}_1,...\hat{K}_k)}
\end{eqnarray}
- is the Jacobian of the transformation of the variables \(H,\;
K_1,...K_k\) into the variables \(H,\; K_1,...K_k\). This Jacobian
is constant on the shared level surface of the functions \(H,\;
K_1,...K_k\), which proves the correctness of the definition.

We will call distribution (\ref{MKdis}) the generalized
micro-canonical distribution. It has to be said that the generalized
micro-canonical distribution can be derived from the requirement
that the Von Neumann entropy must achieve its maximum if the energy
value and the values of the first integrals  \(K_1,...,K_k\), are
fixed. In other words, the generalized micro-canonical distribution
corresponds to a state of the system in which the information that
is available about it (with the fixed values of the Hamiltonian and
the first integrals \(K_i,...,K_k\)  is minimal.

Using a method similar to that we used above, we can demonstrate
that the micro-canonical distribution does not change its form as a
result of quasi-static processes. The following formula has to be
noted here: if the Hamiltonian \(H\) and the first integrals
\(K_1,...,K_k\) are functions of parameter \(\lambda\), then in
quasi stationary processes:
\begin{eqnarray}
(\frac{\partial
K'_i}{\partial\lambda})_{qs}=\overline{\frac{\partial
K_i}{\partial\lambda}},
\end{eqnarray}
where \(qs\) means that the derivative is calculated along the
quasi-static process. We will now define the entropy corresponding
to the micro-canonical distribution as:
\begin{eqnarray}
S(E,K'_1,...,K'_k)=\rm ln \mit W(E,K'_1,...,K'_k),
\end{eqnarray}
Where \(W(E,K'_1,...,K'_n)\) is the so called statistical weight:
\begin{eqnarray}
W(E,K'_1,...,K'_k)=\int d\Gamma_x \delta(H(x)-E)\prod
\limits_{i=1}^k\delta(K_i(x)-K'_i).
\end{eqnarray}
Let's found out how correct this definition is, i.e. how will the
entropy change if we move from the integrals \(H,\;K_1,...,K_k\) to
the integrals \(H,\;\hat{K}_1,...,\hat{K}_k\). What we get is:
\begin{eqnarray}
W(E,K'_1,...,K'_n)=c\int d\Gamma_x \delta(H(x)-E)\prod
\limits_{i=1}^n\delta(\hat{K}_i(x)-\hat{K}'_i),
\end{eqnarray}
where \(c\) is the transformation Jacobian. The appearance of this
multiplier next to the statistical weight will create an additional
term in the entropy \(\rm ln \mit\, c\). It has to be noted here,
that a similar problem also arises in standard thermodynamics (\(k =
0\)), where this parameter is simply discarded because it is assumed
that it's influence in the thermodynamic limit is infinitesimally
small. We will do the same here. It also has to be noted that
instead of (\ref{MKdis}) the following formula can be used for the
micro-canonical distribution:
\begin{eqnarray}
\rho(x)=c\Delta(H(x)-E)\prod \limits_{i=1}^k\Delta(K_i(x)-K'_i),
\label{Delta}
\end{eqnarray}
Where function \(\Delta\) equals one in the small neighborhood of
zero and equals zero outside this neighborhood. In the thermodynamic
limit, this neighborhood will be so narrow that \(\Delta\)-function
can be replace with \(\delta\)-function, and we will get back to
(\ref{MKdis}). In light of the above notes, the statistical weight
can be defined as:
\begin{eqnarray}
W(E,K'_1,...,K'_n)=\int d\Gamma_x \Delta(H(x)-E)\prod
\limits_{i=1}^k\Delta({K}_i(x)-{K}'_i).
\end{eqnarray}
It can be seen from this formula that the statistical weight can be
interpreted as the number of microscopic states compatible with a
given macroscopic state, or as the thermodynamic probability.

Now let's assume our system is in equilibrium with the environment.
Let's demonstrate that we can choose the first integrals
\(\hat{K}_1,...,\hat{K}_n\) in such a way that their values will
remain the same as the system heats up. And indeed, the energy level
E corresponds to the integral values  \(K'_1(E),...,K'_n(E)\). Let's
replace the integrals \(K_1,...,K_n\) with the integrals
\(\hat{K}_1,...,\hat{K}_n\) using the formula
\(\hat{K}_i=K_i-K'_i(H)\). It can be seen that these integrals
remain constant as the system heats up. We then assume that these
integrals also remain constant when the system temperature
increases. What we then have is
\begin{eqnarray}
(\frac{\partial
K'_i}{\partial\lambda})_{qs}=\overline{\frac{\partial
K'_i}{\partial\lambda}}.
\end{eqnarray}
From which we can derive: But because:
\begin{eqnarray}
(\frac{\partial K'_i}{\partial\lambda})_{qs}=(\frac{\partial
K'_i}{\partial\lambda})_{E}+(\frac{\partial K'_i}{\partial
E})_\lambda (\frac{\partial E}{\partial\lambda})_{qs},
\end{eqnarray}
and \((\frac{\partial K'_i}{\partial E})_\lambda=0\) we have
\begin{eqnarray}
(\frac{\partial K'_i}{\partial\lambda})_{E}=(\frac{\partial
K'_i}{\partial\lambda})_{qs}=\overline{\frac{\partial
K_i}{\partial\lambda}}. \label{74}
\end{eqnarray}
It has to be noted that the entropy remains constant in a
quasi-static process. Indeed, in a quasi-static process, the
(generalized) micro-canonical distributions corresponding to the
various parameters turn out to be bound, as was demonstrated above,
by a canonical transformation and their entropy is the same because
canonical transformations preserve the phase volume. One important
result of this section is the conclusion that in the thermodynamics
of stationary non-equilibrium states, the standard thermodynamic
equations hold true. Let's derive the standard thermodynamic
equation:
\begin{eqnarray}
dE=TdS-PdV, \label{TD}
\end{eqnarray}
where \(T\) is the temperature, \(P\) is the pressure, and \(V\) is
the volume of the system. Put by definition that
\begin{eqnarray}
\frac{1}{T'}=\frac{dS}{dE}
\end{eqnarray}
and demonstrate that \(T' = T\),  i.e. that it is the same as the
ordinary temperature. Let \(E'\) be the energy of the environment
around the system and \(S(E')\) its entropy. We will  assume that
our system is in equilibrium with its environment (but it's not in
equilibrium with itself, being in a stationary non-equilibrium
state.) The probability \(P(E, E')\) that the system's energy equals
E, and the energy of the environment equals  \(\sim
e^{S(E)+S'(E')}\) because we earlier defined the system's entropy
\(S(E)\) as the logarithm of thermodynamic probability. If the
system is in equilibrium with its environment, then  \(P(E, E')\)
reaches its maximum (with an additional condition that \(E+E'=\rm
const \mit\)), because equilibrium is the most probable state in
this case. Consequently:
\begin{eqnarray}
(\frac{dS(E)}{dE}-\frac{dS'(E')}{dE'})dE=0,
\end{eqnarray}
i.e.
\begin{eqnarray}
T'=T,
\end{eqnarray}
and that is exactly what we've been trying to prove.

Let us remind you that \(S = \ln W\), where
\begin{eqnarray}
W=\int d\Gamma_x\delta(E-H(x))\prod
\limits_{i=1}^n\delta(K'_i(V)-K_i(x)).
\end{eqnarray}
Here we directly specify the dependence of \(K_i\) on \(V\). We just
demonstrated that \((\frac{\partial S}{\partial E})_V=\frac{1}{T}\).
Now all we've got left to do is to demonstrate that
\((\frac{\partial S}{\partial V})_E=\frac{1}{T}P\). Hereinafter
derivatives are taken with  \(V = V_0\) for some \(V_0\). What we
get is:
\begin{eqnarray}
(\frac{\partial S}{\partial V})_E=\nonumber\\
=\frac{1}{W} \int d \Gamma_x \{-\frac{\partial H(x)}{\partial
V}\delta'(E-H(x))\prod
\limits_{i=1}^{n}\delta(K'_i(V)-K(x))+\nonumber\\
+\delta(E-H(x))\sum \limits_{j=1}^n [\prod \limits_{i=1,\;i\neq
j}\delta(K'_i-K_i(x))\delta'(K'_j-K_j(x))(\frac{\partial
K'_j(V)}{\partial V}-\frac{\partial K_j(x)}{\partial V})]=\nonumber\\
=\frac{1}{W} \int d \Gamma_x \{-\frac{\partial H(x)}{\partial
V}\delta'(E-H(x))\prod
\limits_{i=1}^{n}\delta(K'_i-K(x))+\nonumber\\
+\sum \limits_{j=1}^n \frac{1}{W}\frac{\partial}{\partial
K'_j}[W((\frac{\partial K'_j}{\partial
V})_E-\overline{\frac{\partial K_j(x)}{\partial
V}})]\}-\frac{\partial}{\partial V}(\frac{\partial K'_j(V)}{\partial
K'_j(V_0)})\label{80}.
\end{eqnarray}
Because of equation (\ref{74}) the second to last term on the right
of the last equation  (\ref{80})  equals zero. As far as the last
term is concerned, the most reasonable assumption would be that
\(\frac{\partial K'_j(V)}{\partial K'_j(V_0)}\approx 1\)  or
\(\frac{\partial K'_j(V)}{\partial K'_j(V_0)}=1-\rm const \mit
\frac{V-V_0}{V_0}\), with \(V\approx V_0\). Therefore the last term
in (81) disappears in the thermodynamic limit. Thus the final result
is:
\begin{eqnarray}
(\frac{\partial S}{\partial V})_E=-\frac{1}{W} \int d \Gamma_x
\frac{\partial H(x)}{\partial V}\delta'(E-H(x))\prod
\limits_{i=1}^{n}\delta(K'_i-K(x))=\nonumber\\
=-\frac{1}{W} \frac{\partial}{\partial E} \int d \Gamma_x
\frac{\partial H(x)}{\partial V}\delta(E-H(x))\prod
\limits_{i=1}^{n}\delta(K'_i-K(x))=\nonumber\\
=-\frac{1}{W}\frac{\partial}{\partial E} (\overline{(\frac{\partial
H(x)}{\partial V})}W).
\end{eqnarray}
But \(-\overline{(\frac{\partial H(x)}{\partial V})}\), as was
already demonstrated earlier, is the speed at which the energy
changes in an adiabatic process, in which the changing parameter is
the volume, i.e. pressure. Thus
\begin{eqnarray}
(\frac{\partial S}{\partial
V})_E=\frac{1}{W}\frac{\partial}{\partial E}(PW)=\nonumber\\
=P\frac{\partial S}{\partial E}+\frac{\partial P}{\partial E}.
\end{eqnarray}
But energy is an extensive value while pressure is an intensive one.
For this reason in the thermodynamic limit \(\frac{\partial
P}{\partial E}=0\). Finally, remember that \(\frac{\partial
S}{\partial E}=\frac{1}{T}\) we get:
\begin{eqnarray}
(\frac{\partial S}{\partial V})_E=\frac{1}{T}P,
\end{eqnarray}
This is what we've been trying to prove. U

sing the standard method and applying the Legendre transformation we
can introduce thermodynamic functions. The free energy  F is defined
as:
\begin{eqnarray}
F:=E-TS.
\end{eqnarray}
Using (\ref{TD}) we can find:
\begin{eqnarray}
dF=-SdT-PdV.
\end{eqnarray}
Similarly we can define the thermodynamic potential  \(\Phi\):
\begin{eqnarray}
\Phi:=F+PV.
\end{eqnarray}
What we get is:
\begin{eqnarray}
d\Phi=-SdT+VdP.
\end{eqnarray}
If the number of particles in the system \(N\) is variable then
(\ref{TD}) should be changed to:
\begin{eqnarray}
dE=TdS-PdV+\mu dN.
\end{eqnarray}
Using the theorem of small additions we can find that:
\begin{eqnarray}
dF=-SdT-PdV+\mu dN,\nonumber\\
d\Phi=-SdT+VdP+\mu dN.
\end{eqnarray}
Integrating the last equation with constant \(V\) and \(T\) between
zero and \(N\), we can find that:
\begin{eqnarray}
\Phi=\mu N.
\end{eqnarray}
If the system has particles of various sorts numbered with \(i =
1,..., M\) and \(N_i\) is the number of particles of sort \(i\),
then in all the previous formulas the term  \(\mu d N\) needs to be
replaced with \(\sum \limits_{i=1}^M \mu_i dN_i\). We also have:
\begin{eqnarray}
\Phi=\sum \limits_{i=1}^M\mu_i N_i.
\end{eqnarray}
In all of the above all the thermodynamic values, including the
entropy, proved to be dependent on the values \(K'_1,...,K'_k\) of
the motion integrals \(K_1,...,K_k\). Now we're going to formulate
the following

\textbf{Equivalence principle.} The entropy \(S(E,K'_1,...,K'_k)\)
does not depend on \(K'_1,...,K'_k\) (if all the other parameters of
the system are constant).

Here is how it can be substantiated. We assigned some specific value
to the energy \(E\), and for the sake of simplicity assume that
\(n=1\). We normalize \(K':=K'_1\) in such a way as to ensure that
\(K'\) changes from \(-1\) to \(1\) and that at \(0\) the entropy
reaches its maximum.

If \(S(K')\) depends on \(K'\) and our system is in thermodynamic
equilibrium with its environment then it will be established such a
value of \(K'\) that corresponds to maximum entropy, i.e. \(K' =
0\).

In a sufficiently small neighborhood of \(K' = 0\), \(S(K')\) will
look like this:
\begin{eqnarray}
S(K')=S_0-c{K'}^2,
\end{eqnarray}
\(c> 0\). Because our system is macroscopic and \(c\) must be
comparable to \(S_0\), we have to assume that \(c\) is very great.
In this case the thermodynamic probability will look like this:
\begin{eqnarray}
W(K')=\rm const \mit e^{-c{K'}^2},
\end{eqnarray}
i.e. \(W(K')\) will have a pronounced peak at \(K' = 0\), while all
the other values of \(K'\) will be extremely improbable. This means
that, in other words, that it doesn't really matter whether we're
averaging using the micro-canonical distribution
\begin{eqnarray}
\rho(x)=c\delta(E-H(x))
\end{eqnarray}
or using the generalized micro-canonical distribution:
\begin{eqnarray}
\rho(x)=c\delta(E-H(x))\delta(K'-K(x)).
\end{eqnarray}
Thus if \(S(E, K')\) depends on \(K'\) then we don't have to use the
micro-canonical distribution  with a fixed value of the integral
\(K\).

Thus the most general case which we can come across is where the
entropy \(S(H,K'_1,...,K'_k)\) depends only on the values of the
integrals \(K_1,...,K_l\), \(l=1,...,k\)  and does not at all depend
on the values of the integrals \(K_{l+1},...,K_k\). The integrals
\(K_1,...,K_l\) are assigned values corresponding to maximum entropy
and the mean for the generalized micro-canonical distribution,
corresponding to the values of the integrals
\(K_1=K'_1\),..\(K_l=K'_l\), \(K_{l+1}=K'_{l+1}\),...,\(K_k=K'_k\)
will equal the meant for the distribution:
\begin{eqnarray}
c\delta(H-E)\prod \limits_{i=l+1}^k\delta(K_i-K'_i).
\end{eqnarray}
The observed values of the integrals \(K_{l+1},...,K_k\) can be
arbitrary.

Now let us consider an example from the condensed matter physics
where the general case above is realized. The case in hand is the
superfluid helium.

Let us consider a system of \(N\) particles enclosed in a certain
macroscopic volume \(V\) and obeying the Bose statistics. The
Hamiltonian of such system takes the form:
\begin{eqnarray}
H=\sum \limits_{i=1}^N \frac{p_i^2}{2m}+\sum \limits_{i=1}^N \sum
\limits_{j=i+1}^N \Phi(|x_i-x_j|),
\end{eqnarray}
where \(\Phi(|x_i-x_j|)\) is the potential energy of the \(i\)-th
and \(j\)-th particles.

Let us assume that
\begin{eqnarray}
\nu(p)=\int \Phi(|x|)e^{ipx}d^3x.
\end{eqnarray}
Then, in secondary quantization representation the Hamiltonian will
take the form:
\begin{eqnarray}
\Gamma:=H-\mu N=\sum \limits_p
(\frac{p^2}{2m}-\mu)b_p^+b_p+\nonumber\\
+\frac{1}{2V} \sum
\limits_{p_1,p_2,p'_1,p'_2}\Delta(p_1+p_2-p'_1-p'_2)\nu(p_1-p_1')b^+_{p_1}b^+_{p_2}b_{p'_1}b_{p'_2}.
\end{eqnarray}
Here \(\Delta(x)\) is a the function equal to 1 when \(x=0\) and
equal to zero in all the other cases. \(b_p^+\), \(b_p\) are
particle creation-annihilation operators in a state with the
momentum of \(p\).

With low interaction between the particles almost all of them are in
condensed state. And as \(b_0,\;b_0^+\sim \sqrt{N}\), in commuting
relation we can neglect 1 and consider \(b_0\), \(b_0^+\)
commutation variables. Let us denote the number of particles in the
condensate by \(N_0\).

Keeping only the terms that are quadratic by the
creation-annihilation operators of the super-condensate particles in
the Hamiltonian, we derive:
\begin{eqnarray}
\Gamma=-\frac{1}{2}\frac{N_0^2}{V} \nu(0)+\sum \limits_{p\neq0}
\{\frac{p^2}{2m}+\frac{N_0}{V} \nu(p)\}b_p^+b_p+\nonumber\\
+\frac{N_0}{2V}\sum \limits_{p\neq0}\nu(p)(b^+_pb^+_{-p}+b_{-p}b_p).
\end{eqnarray}

This Hamiltonian can be diagonalized by the canonical Bogoliubov
transformation [13]:
\begin{eqnarray}
\xi_p:=\frac{b_p-A_pb^+_{-p}}{\sqrt{1-A_p^2}},\nonumber\\
\xi_p^+:=\frac{b_p^+-A_pb_{-p}}{\sqrt{1-A_p^2}},
\end{eqnarray}
where
\begin{eqnarray}
A_p=\frac{V}{N_0\nu(p)}\{E(p)-\frac{p^2}{2m}-\frac{N_0}{V}\nu(p)\},\nonumber\\
E(p)=\sqrt{\frac{N_0}{V}\frac{p^2\nu(p)}{m}+\frac{p^4}{4m}}.
\end{eqnarray}
In new variables the Hamiltonian of the system will take the form:
\begin{eqnarray}
H=H_0+\sum \limits_{p\neq0} E_p\xi^+_p\xi_p.
\end{eqnarray}
If the condensate moves at non-zero rate \(u\), then the Hamiltonian
of the systems takes the form:
\begin{eqnarray}
H=H_0+\sum \limits_{p\neq0} (E_p-up)\xi^+_p\xi_p.
\end{eqnarray}
This Hamiltonian has the evident motion integrals --- the occupation
numbers. Or rather let us separate the set of all non-zero momentums
\(\{p\}\) to disjoint subsets \(S_i\), containing \(G_i\) elements,
corresponding to the close values of the momentums. We assume that
\(G_i\rightarrow \infty\) when \(V\rightarrow \infty\), but,
together with that, \(\frac{G_i}{V}\rightarrow 0\) when
\(V\rightarrow \infty\). Then the motion integrals will be the
occupation numbers \(N_i\) of the cells \(S_i\). Let us assume
\(n_i=\frac{N_i}{G_i}\). The entropy of the system will take form:
\begin{eqnarray}
S=\sum \limits_i G_i[(1+n_i)\ln (1+n_i)-n_i \ln n_i].
\end{eqnarray}
This value reaches the maximum under the fixed energy and number of
particles on Bose distribution.

On the other side, except the occupation numbers, Bose gas has one
more first integral --- the condensate motion velocity. Let us
analyze the dependence of the thermodynamic values of our system on
the value of this integral. The state of the liquid helium is
convenient to be described via the condensate wave function \(\Xi\)
[14]. It satisfies the following differential equation:
\begin{eqnarray}
i\frac{\partial}{\partial
t}\Xi(x,t)=-\frac{1}{2m}\nabla^2\Xi(x,t)+U_0\Xi(x,t)|\Xi(x,t)|^2-c\Xi(x,t),
\end{eqnarray}
where \(U_0,\;c >0\).

If the system is kept at a constant temperature and its state is
described by the number of \(\lambda_1,...,\lambda_l\), then in
equilibrium the free energy \(F(T,\lambda_1,...,\lambda_l)\) must
reach the minimum. In our case  the condensate wave function
\(\Xi(x,t)\) performs the role of these parameters. But the
condensate wave function can vary with time and in meaning,
\(\Xi(x,t),\;\Xi^+(x,t)\) represent canonically conjugate variables.
That is why the requirement of free energy minimality should be
replaced by the requirement of stationary action:
\begin{eqnarray}
A[\Xi,\Xi^+]=\int \limits_{t_1}^{t_2}[i \int d^3
x\Xi^+\frac{\partial}{\partial t}\Xi-F[\Xi,\Xi^+]]dt.
\end{eqnarray}
As \(F\) is defined accurate to constant, we will find:
\begin{eqnarray}
F[\Xi,\Xi^+]=\int d^3x \{-\frac{1}{2m}
\Xi^+(x)\nabla^2\Xi(x)+\frac{U_0}{2}|\Xi(x)|^4-c|\Xi(x)|^2\}.
\end{eqnarray}
To show that the thermodynamic characteristics do not depend on the
velocity of the condensate motion it is sufficient to show that the
action \(A[\Xi,\Xi^+]\) is invariant under Galilean transformation:
\begin{eqnarray}
\Xi(x,t)\mapsto e^{i(\frac{mv^2}{2}t+mvx)}\Xi(x-vt,t),
\end{eqnarray}
\(v\) is relative motion velocity. But the invariance of the action
\(A[\Xi,\Xi^+]\) in relation to the Galilean transformation is easy
to find out by direct calculation.

So, the action \(A[\Xi,\Xi^+]\) does not depend on the condensate
flow velocity and this fact provides the stability of the condensate
flow.

We needed this example to illustrate the stability of the
generalized microcanonical distributions. Besides we obtained the
answer to the question of why not all the first integrals residing
in the system are observed. Only those integrals are observed that
have the entropy unchanged while the values of these integrals are
changing (at constant values of other integrals).

In the next section when we analyze the Ling cell, it will be
essential that some of the first integrals of the Ling cell satisfy
the equivalence principle. In addition to that, these first
integrals have never been observed directly the way that the
velocity of the Bose condensate through the capillary can be
observed. These first integrals could only be observed indirectly by
analyzing physiological phenomena testifying their presence. For
this reason there would be a lot of doubts about or theory if no
inorganic systems could be found for which there exist non-trivial
first integrals satisfying the equivalence principle and which can
therefore be observed. The example cited above allows us to dispense
with such doubts.

Let us discuss the equivalence principle more detailed. For the sake
of simplicity we will assume that the entropy of the system
\(S(E,K')\) depends on the value of \(K'\) of one integral \(K\),
however, the reasoning given below is directly generalized for the
case when the entropy depends on the values of the two and more
first integrals in involution.

According to the equivalence principle the entropy \(S(E,K')\) does
not depend on \(K'\). But it would be more precise to say that the
entropy reaches its maximum on the certain interval of the values of
the integral \(K'\) \([\alpha(\lambda),\beta(\lambda)]\) and steeply
decreases out of this interval. Here \(\lambda\) --- the parameter
on which the Hamiltonian of the system depends. Here the energy of
the system depending on the parameter \(\lambda\) is chosen the way
that at any value of \(\lambda\) the entropies of the system were
always the same.

At the adiabatic infinitesimal change of the parameter \(\lambda\),
\(\lambda_0\mapsto\lambda'=\lambda_0+\delta\lambda\) the value of
the integral \(K'\) is transformed by the formula \(K'\mapsto
K'+\overline{\frac{\partial K}{\partial \lambda}}\delta\lambda.\)
Let us show that at this process the plateau
\([\alpha(\lambda_0),\beta(\lambda_0)]\) transforms into the plateau
\([\alpha(\lambda'),\beta(\lambda')]\).

Our system is in equilibrium with the thermostat and let us assume
that \(K'_e \in [\alpha(\lambda_0),\beta(\lambda_0)]\) any
equilibrium value of the integral \(K'\) (\(\lambda=\lambda_0\)).
Let us show that when the system's energy is fixed at \(E\) and \(K'
\in [\alpha(\lambda_0),\beta(\lambda_0)]\) the temperature of the
system does not depend on \(K'\).

Let us assume that \(T\) --- the temperature corresponding to the
equilibrium value of \(K'_e\) of the integral \(K\), and equal to
the temperature of the thermostat.  Let us assume that \(K'_1\)
--- any value of the integral \(K\) from the interval
\([\alpha(\lambda_0),\beta(\lambda_0)]\). Let us assume that \(T'\)
--- The temperature corresponding to it, and \(T'\neq T\). Let us
surround our system by an adiabatic membrane and change
\(K'_e\mapsto K'_1\). When this happens neither the system energy +
thermostat nor the system entropy + thermostat are not changing. Now
let us make our system contact the thermostat again which will lead
to the equalization of the temperatures of the system and the
thermostat. At this stage we assume that system + thermostat is an
isolated system, particularly, its full energy is constant. As at
the process of temperature equalization a self-induced heat flow
takes place, the entropy of the system and the thermostat increases.
That means, at the value of the integral \(K\) \(K'=K'_e\) for fixed
full value of the common energy of the system and the thermostat,
the entropy of the system and the thermostat is not maximal which
contradicts the fact that \(K'_e\)
--- the equilibrium value of the integral \(K\).

The same way we will demonstrate that at the fixed energy of the
system the pressure \(P_1\), corresponding to the value of the
integral \(K'_1\) Is equal to the ambient pressure \(P\). The
pressure is defined by formula \(P=(\frac{\partial E}{\partial
\lambda})_S\). Let us assume, for definiteness, that \(P_1>P\). As
it was demonstrated above the temperature corresponding to the value
of the integral \(K'_1\) is equal to the ambient temperature \(T\).
Let us make the system “expand” isothermally, doing work against the
external forces. Meanwhile the system pressure \(P_1\) will be
decreasing by virtue of the known thermodynamic inequation, fair in
our case as well, that states that  the module of the isothermal
compressibility is positive. We will allow the system to expand
quasistatically until its pressure is equal to the pressure of the
thermostat. At that the system + thermostat will do a certain
positive work, i.e. the full energy of the system + thermostat will
decrease by this value. But as the process of the system expansion
is quasistatic and therefore reversible, the full entropy of the
system + thermostat will not change. Therefore we can conclude that
at the value of the integral \(K'_e\), and for fixed full entropy of
the system and the thermostat, the energy minimum is not reached
which means the value of the integral \(K'_e\) is not equilibrium.

So, we demonstrated that for fixed energy of the system \(E\), The
values of the pressure and temperature are constant on the whole
interval \([\alpha(\lambda_0),\beta(\lambda_0)]\) of the values of
the integral \(K'\).

Let us now adiabatically change the \(\lambda\),
\(\lambda_0\mapsto\lambda'=\lambda_0+\delta\lambda\). At this change
the entropy of the system will not change and the energy of the
system will increase by the value \(P\delta\lambda\), i.e. the value
that does not depend on \(K'\). In other words, at the
transformation \(K'\mapsto K'+\overline{\frac{\partial K}{\partial
\lambda}}\delta\lambda\) the points of the plateau will pass again
into points of the plateau but corresponding to the changed
parameter \(\lambda\). When making the inverse adiabatic change of
the parameter \(\lambda\)
\(\lambda'\mapsto\lambda_0=\lambda'-\delta\lambda\), we see that the
formula \(K'\mapsto K'+\overline{\frac{\partial K}{\partial
\lambda}}\delta\lambda\) sets a one-to-one correspondence between
the points of the plateau \([\alpha(\lambda_0),\beta(\lambda_0)]\)
and \([\alpha(\lambda'),\beta(\lambda')]\).

We remind that we proceed from the assumption that the system is in
equilibrium with the environment but is not in equilibrium with
itself, being in a stationary non-equilibrium state. Let us describe
a possible way for the system to change from stationary
non-equilibrium to equilibrium. Let us use the (\ref{Delta}) for the
distribution function. This expression uses \(k\) independent first
integrals in involution \(K_1,...,K_k\). However that does not mean
that there no other integrals in the system except \(K_1,...,K_k\),
being in involution with them. For example in the formula for the
ordinary microcanonical distribution only the energy function takes
part but, as it was discussed in the section 1, even the most
realistic systems of the statistical mechanics are nonergodic at the
thermodinamyc limit, i.e. there are other first integrals except the
energy. We will call first integrals that are included in formula
(\ref{Delta} active. A possible way of changing to equilibrium is
that the more and more first integrals from the list \(K_1,...,K_k\)
cease being active.

Let us assume that in this system the active integrals were
\(K_1,...,K_k\) and the integral \(K_k\) ceased being active. Let us
use the symbol \((\delta S)_E\) to denote the change of the entropy
at this process (at constant energy). Let us prove the following
important proposition:

\textbf{Proposition.} \((\delta S)_E\) does not change at
quasistatic processes, that is to say
\begin{eqnarray}
(d(\delta S)_E)_{qs}=0.
\end{eqnarray}

\textbf{Proof.} Let us assume that the active integrals are
\(K_1,...,K_k\). Then the generalized microcanonical distribution
corresponding to these integrals will take the form:
\begin{eqnarray}
w_k(E,K'_1,...,K'_k,x)=c\Delta(H(x)-E) \prod
\limits_{i=1}^k\Delta(K_i(x)-K'_i).
\end{eqnarray}
Let us note that \(w_k(E,K'_1,...,K'_k,x)\) is in proportion to the
characteristic function of a certain set. If the integral \(K_k\)
becomes inactive then the generalized microcanonical distribution oi
\(w_{k-1}(E,K'_1,...,K'_{k-1},x)\) turns out to be composed of a
certain number \(M\) of microcanonical distributions
\(w_k(E,K'_1,...,K'_k,x)\), the entropies of which are the same. It
is evident that \((\delta S)_E=\ln M\).

Let us consider the adiabatic change of the parameter
\(\lambda\mapsto\lambda'=\lambda_0+\delta\lambda\). We can assume
that \(\forall i=1,...,k\)
\begin{eqnarray}
\overline{(\frac{\partial K_i}{\partial
\lambda})}_{K'_1,...,K'_k}|_{\lambda=\lambda_0}=0.
\end{eqnarray}

In this formula the low indices \(K'_1,...,K'_k\) mean that the
averaging corresponding to the vinculum is taken over the
microcanonical distribution \(w_k(E,K'_1,...,K'_k,x)\). This can be
reached by replacing \(\forall i=1,...,k\) \(K_i\mapsto
K_i-(\lambda-\lambda_0)f_i(K_1,...,K_k,H)\), where
\(f_i(K_1,...,K_k, H)\) is the suitable functions of the first
integrals and Hamiltonian.

We will suppose that at this adiabatic process the point in the
phase space that represents the state of our system, constantly
moves in such a way that the values \(K_1,...,K_k\) are the motion
integrals. Meanwhile the microcanonical distribution
\(w_k(E,K'_1,...,K'_k,x)\) transforms into the microcanonical
distribution \(w_k(E',K'_1,...,K'_k,x)\). But, because of the above,
the value \(\delta E:=E'-E\) will not depend on the value \(K'_k\)
and will be the same if we calculat it for the distribution
\(w_{k-1}(E,K'_1,...,K'_{k-1},x)\). In the same adiabatic process
the microcanonical distribution \(w_{k-1}(E,K'_1,...,K'_{k-1},x)\),
corresponding to the value of the parameter \(\lambda=\lambda_0\),
will transform into the distribution
\(w_{k-1}(E',K'_1,...,K'_{k-1},x)\), and by virtue of the fact that
\(\delta E\) can be calculated over the distribution
\(w_{k-1}(E,K'_1,...,K'_{k-1},x)\), the entropies corresponding to
the distribution \(w_{k-1}(E,K'_1,...,K'_{k-1},x)\) with
\(\lambda=\lambda_0\) and the distribution
\(w_{k-1}(E',K'_1,...,K'_{k-1},x)\) with \(\lambda=\lambda'\) will
be the same. But at the adiabatic process (such as the one described
in the beginning of the proof) the evolution of the distributions
simply resolves into the canonical substitution of the arguments of
these distributions. That is why with \(\lambda=\lambda'\), the
distribution \(w_{k-1}(E',K'_1,...,K'_{k-1},x)\) will turn out to be
composed from the same number of \(M\) generalized microcanonical
distributions \(w_k(Ey,K'_1,...,K'_k,x)\). That is to say \(M=\rm
const \mit\). That is to say \(((\delta S)_E)_{qs}=\ln M=\rm const
\mit\), which was to be proved.

In case if there are too many of first integrals and if the process
of decrease of the number of active integrals can be considered
continuous, it is reasonable to introduce the continuous parameter
\(s\), characterizing the number of active integrals the way that
\(s \in [0,1]\), the value \(s=0\) corresponds to the case when no
integrals are active and the decrease of the number of active
integrals corresponds to the decrease of the parameter \(s\).
Therefore we have:
\begin{eqnarray}
\frac{dS}{ds}\leq 0.
\end{eqnarray}

The formula \(\frac{dS}{ds}\leq 0\) is correct when it is assumed
that the system energy and volume are kept constant. Let us assume,
however, that the system temperature and volume are kept constant.
Then:
\begin{eqnarray}
0 \geq(\frac{dS}{ds})_{E,V}=(\frac{dS}{ds})_{T,V}-(\frac{\partial
S}{\partial E})_{V,s}(\frac{dE}{ds})_{V,T}=\nonumber\\
=(\frac{dS}{ds})_{T,V}-\frac{1}{T}(\frac{dE}{ds})_{V,T}=\nonumber\\
=\frac{1}{T}(\frac{d(ST-E)}{ds})_{T,V}.
\end{eqnarray}
That is to say:
\begin{eqnarray}
(\frac{dF}{ds})_{T,V}\leq 0.
\end{eqnarray}
The same way let us assume that the system temperature and pressure
are kept constant. Then:
\begin{eqnarray}
0\geq (\frac{d F}{d s})_{T,V}=(\frac{d F}{d
s})_{T,P}-(\frac{\partial F}{\partial V})_{s,T}(\frac{d V}{d
s})_{P,T}=\nonumber\\
=(\frac{d F}{d s})_{T,P}-P(\frac{d V}{d s})_{P,T}=\nonumber\\
=(\frac{d \Phi}{d s})_{P,T}.
\end{eqnarray}

So, in case when the temperature and pressure are kept constant:
\begin{eqnarray}
(\frac{d \Phi}{d s})_{P,T}\leq 0.
\end{eqnarray}

The statement, that the number of independent first integrals in
involution for the system must be very large can be established, for
example, by the following way. Let us decompose the system
\(\mathfrak{S}\) into disjoint union of enough large number \(N\) of
small but macroscopic subsystems \(\mathfrak{S}_i,\;i=1,...,N\). Let
\(H_i,\;i=1,...,N\) be a Hamiltonian of subsystems
\(\mathfrak{S}_i\) and \(H_{ij},\;i,j=1,...,N,\;i\neq j\) be the
total interaction Hamiltonian of subsystems \(\mathfrak{S}_i\) and
\(\mathfrak{S}_j\). Then \(\sum \limits_{i=1}^N H_i\) is
proportional to the volume of system \(\mathfrak{S}\) and \(\sum
\limits_{1\leq i<j\leq N}^N H_{i,j}\) is proportional to the total
area of boundaries between different subsystems \(\mathfrak{S}_i\).
But each subsystem \(\mathfrak{S}_i\) is macroscopic. So we can
neglect by the interaction Hamiltonian of different subsystems in
total hamiltonian \(H\) of system \(\mathfrak{S}\). Therefore we can
write that
\begin{eqnarray}
 H=\sum\limits_{i=1}^N H_i \label{additiv}
\end{eqnarray}
But each subsystem \(\mathfrak{S}_i\) is macroscopic. So, according
to the nonergodic theorem for each subsystem \(\mathfrak{S}_i\)
there exists at least one its first integral \(K_i\). It is easy to
see that for different \(i=1,...,N\) integrals \(K_i\) are
independent and in involution. It follows from (\ref{additiv}) that
integrals \(K_i\) are the first integrals of whole system
\(\mathfrak{S}\) at the same time.

\section{ E.S. Bauer's stable non-equilibrium principle and the resting state of G. Ling.}

For the sake of further analysis we will consider the contribution
that E.S. Bauer and G. Ling made to the issue at hand. According to
Bauer [4] the substance of the living cell can be in one of two
states: the stable-non-equilibrium (resting) and thermodynamic
equilibrium (active). The work The work produced by living system a
living system is done when the system passes from stable
non-equilibrium to equilibrium. Bauer used this principle (the two
basic states and the change between them) as the basis for his
"theoretical biology" and deductively derived the basic properties
of such biological processes as assimilation, growth, excitability,
adaptability and reproduction. That is what Bauer himself says [4,
p. 143] about this principle:

"Nonliving systems are ever in equilibrium and due to their free
energy they do work against equilibrium required by the laws of
physics and chemistry in the existing ambient conditions."Further
Bauer expands on it: "We will call this principle as stable
none-equilibrium principle of the living systems. This name clearly
expresses the meaning of the principle and the unique feature of
living systems from the point of view of thermodynamics. The same
way as a system in stable equilibrium when disturbed returns back to
it, a living system in stable non-equilibrium also keeps coming back
to it. Our principle also characterises the unique feature of living
systems as we do not know any nonliving systems that would be in
stable non-equilibrium."

The ground rule of Ling's Association-Induction Hypothesis [1,2,3]
is that the cell as a quasi-solid body is a system with a
non-maximal entropy in a resting state.We proceed from the approach
based on the fact that the resting state of the living cell,
according to both Ling and Bauer, is a stationary non-equilibrium
state, the possibility of existence of which was proven by one of us
for a large class of realistic systems in statistical mechanics [7].
Here we simply apply it to the living cell. The appropriateness of
the proposed approach to the living will be verified by comparing
the results of the theoretical analysis and the available
experimental data.

Based on the assumption that the state of the living cell is
stationary non-equilibrium we will give a thermodynamic description
of the following phenomena:

1)  When the cell is excited and when it dies heat is generated.

2) When the cell is excited and when it dies the cells' size changes
(mostly decreases).

3)  When the cell is excited and when it dies , the cell's key
protein molecules fold.

4) When the cell is excited and when it dies, efflux of potassium
ions from the cell takes place.

Let us begin with the explanation of the first phenomenon. Let us
assume that the substance inside the cell is in stationary
non-equilibrium corresponding to a number of active first integrals
in involution and the process of excitation and death manifests
through some of these first integrals becoming inactive. Let's
assume that just like in the section 3 that the number of the active
first integrals is characterised by a continuous \(s \in [0,1]\),
Where \(s = 0\) corresponds to the case, when neither of first
integrals are active and the value of \(s = 1\) to the case when the
maximum number of the first integrals are active. The question why
there must be so many nontrivial integrals that their number can be
described by a continuous parameter, can be answered by the fact
that the muscle fibres can be contracted to different degrees and is
able to algebraically summate the nerve impulses that enter it. It
will be clear when we consider the changes in the size of the Ling
cell when it's activated and when it dies. We can choose \(s\) in
such a way that \(s\) is proportionate to the number of active first
integrals. The system's volume is supposed to be constant.

Let us make an infinitesimal change in the parameter \(s\):
\begin{eqnarray}
s\mapsto s'=s-\delta s,
\end{eqnarray}
where  \(\delta >0\), \(\delta s\) is are infinitely small.

It has been established that the knowledge of \(k\) of the first
integrals in involution allows decreasing the number of degrees of
freedom by \(k\) units (system deflation according to Whit-taker).
For each \(s\) we have \(\rm const\, \mit s\) of active first
integrals and the entropy that corresponds to the parameter \(s\) is
the entropy of the reduced system, corresponding to these constant
\(s\) first integrals. If s  \(s\mapsto s'=s-\delta s\), then the
number of degrees of freedom of reduced system increases by  \(\rm
const \mit \delta s\) units, and we may consider that these new
'turned-on' degrees of freedom corresponding to the new system
\(\delta \mathfrak{S}\), that is in thermodynamic equilibrium with
the initial system  \(\mathfrak{S}\). We will change the parameter
\(s\) at a fixed temperature \(T\). If \(s\mapsto s'=s-\delta s\)
the system's energy changes:
\begin{eqnarray}
E\mapsto E'=E+(\delta E)_T,
\end{eqnarray}
and entropy:
\begin{eqnarray}
S\mapsto S'=S+(\delta S)_T.
\end{eqnarray}
We can equate  \((\delta E)_T\) to the energy of the system \(\delta
\mathfrak{S}\), and \((\delta S)_T\) to the entropy of the same
system. As the system  \(\delta \mathfrak{S}\) is in thermodynamic
equilibrium with the system  \(\mathfrak{S}\), it is essential that
the temperature of the system  \(\delta \mathfrak{S}\) should also
be equal to \(T\), in other words, it is necessary that:
\begin{eqnarray}
\frac{1}{T}=\frac{d (\delta S)_T}{d (\delta E)_T}.\label{RR}
\end{eqnarray}
But this equality can also be derived directly. We have:
\begin{eqnarray}
\frac{dS}{dE}=\frac{1}{T},\nonumber\\
\frac{dS+d (\delta S)_T}{dE+d (\delta E)_T}=\frac{1}{T}.
\end{eqnarray}

From these two equalities we derive  (\ref{RR}) by using the
elementary transformation.

But in order for \(\delta \mathfrak{S}\) to be in equilibrium with
\( \mathfrak{S}\) the equation (\ref{RR} is not enough. It is
necessary that
\begin{eqnarray}
  \frac{d^2 (\delta S)_T}{(d (\delta E)_T)^2}<0. \label{TDuneq}
\end{eqnarray}
This inequality results from the requirement for maximum entropy in
the system  \( \mathfrak{S}+\delta \mathfrak{S}\) under condition
that the energy remains constant, if we take into account the fact
that the number of degrees of freedom of the system  \(\delta
\mathfrak{S}\) is far less than the number of degrees of freedom of
the system \(\mathfrak{S}\). The detailed discussion of this
inequality is found in Appendix 1. This inequality leads to
\begin{eqnarray}
\frac{d}{d (\delta E)_T} (\frac{1}{T})<0,\nonumber\\
-\frac{1}{T^2} \frac{d T}{d (\delta E)_T}<0,
\end{eqnarray}
or
\begin{eqnarray}
\delta c_v:=\frac{d (\delta E)_T}{dT}>0.
\end{eqnarray}
We will make the following assumption about the value of  \( (\delta
E)_T\). There is a temperature \(T_0\), that makes the temperature
\(T> T0\) incompatible with the existence of life in the cell. This
means that if the temperature \(T\rightarrow T_0\)  the interval
\(\Delta\) from the formula (\ref{Delta}) becomes more and more
comparable to the range spaces of the integrals  \(K_1,...K_k\)
values, and the generalized micro-canonical distribution degenerates
into a usual micro-canonical distribution. Then, if \(T > T_0\)
\begin{eqnarray}
(\delta S)_E=\rm const \mit, \label{Gran}
\end{eqnarray}) i.e. if \(T> T_0\) the dependence of \(T\) on \(E\) for \(\mathfrak{S}\) will be the same as the one for
\(\mathfrak{S}+\delta \mathfrak{S}\), i.e. \((\delta E)_T=0\). This
boundary condition is discussed in more detail in the Appendix 2. We
have:
\begin{eqnarray}
  (\delta E)_T=(\delta E)_{T_0}-\int \limits_{T}^{T_0} \frac{d (\delta
  E)_{T'}}{dT'} dT'=\nonumber\\
  = (\delta E)_{T_0}-\int \limits_{T}^{T_0}\delta c_v(T')
  dT'=\nonumber\\
  =-\int \limits_{T}^{T_0}\delta c_v(T')
  dT'<0.
\end{eqnarray}
I.e.  \((\delta E)_T<0\), and that's what we set out to prove
initially.

Let us proceed now to the explanation of why the Ling cell changes
size when it is excited and when it dies. Let us examine a resting
cell in equilibrium with the environment at a constant temperature
\(T\) and pressure \(P\), and show that when the cell is excited and
when it dies its size decreases. So, we assume that as above
\(s\mapsto s'=s-\delta s\), \(\delta s>0\), and that \((\delta
S)_E\) means the change in the entropy of the cell if its energy is
constant.

In section 3 we demonstrated that  \((\delta S)_E=\rm const \mit\)
in an adiabatic process and that is why \((\delta S)_E\) is a
function of the entropy only, that is to say:
\begin{eqnarray}
(\delta S)_E=\delta f(S),
\end{eqnarray}
for some function  \(\delta f\) of one variable. Let us show now
that  \(\delta f\) is a decreasing function. We have:
\begin{eqnarray}
(\delta S)_E=(\delta S)_T-(\delta E)_T \frac{\partial S}{\partial
E},\nonumber\\
(\delta S)_E=(\delta S)_T-\frac{1}{T}(\delta E)_T.
\end{eqnarray}
Because of the thermodynamic inequality  \(c_V>0\)  ((\(c_V\) -the
thermal capacity of the cell at constant volume) the energy, and as
a consequence, the entropy are strictly increasing functions of the
temperature. That is why it is enough to show that
\(\frac{\partial}{\partial T} ((\delta S)_E)<0\). But
\begin{eqnarray}
\frac{\partial}{\partial T} ((\delta S)_E)=\frac{\partial}{\partial
T} ((\delta S)_T)-\frac{1}{T}\frac{\partial}{\partial T} ((\delta
E)_T)+\frac{1}{T^2}(\delta E)_T=\nonumber\\
=\frac{1}{T^2}(\delta E)_T.
\end{eqnarray}
And by virtue of this, as it was already proved, \((\delta E)_T<0\)
we find: \(\frac{\partial}{\partial T} ((\delta S)_E)<0\), i.e.
\(\delta f\) is a  decreasing function of its argument.

Now let us define the minor additions that appear in the
thermodynamic potentials under \(s\mapsto s'=s-\delta s\). We have:
\begin{eqnarray}
(\delta S)_E=(\delta S)_T-(\delta E)_T \frac{\partial S}{\partial
E},\nonumber\\
(\delta S)_E=(\delta S)_T-\frac{1}{T}(\delta E)_T.
\end{eqnarray}

I.e. under  \(s\mapsto s'=s-\delta s\),

\begin{eqnarray}
E(V,S)\mapsto E(V,S)-T\delta f(S).
\end{eqnarray}
According to the minor additions theorem for free energy  \(F\) and
the thermodynamic potential  \(\Phi\) we have:
\begin{eqnarray}
F(V,T)\mapsto F(V,T)-T\delta f(S),\nonumber\\
\Phi(P,T)\mapsto\Phi(P,T)-T\delta f(S).
\end{eqnarray}
As
\begin{eqnarray}
V=(\frac{\partial \Phi(P,T)}{\partial P})_T,
\end{eqnarray}
Then change in the cell size with \(s\mapsto s'=s-\delta s\) takes
the form:
\begin{eqnarray}
\delta V=-T(\delta f)'(S)(\frac{\partial V}{\partial
P})_T(\frac{\partial S}{\partial V})_T.
\end{eqnarray}
As  \((\delta f)'<0\) by virtue of what was proven above and
\((\frac{\partial V}{\partial P})_T<0\) (thermodynamic inequality,
which is always true), the sign \(\delta V\) is opposite to the sign
\((\frac{\partial S}{\partial V})_T\).

 Let us set the sign  \((\frac{\partial S}{\partial V})_T\).
But because of a known thermodynamic relation
\begin{eqnarray}
(\frac{\partial S}{\partial V})_T=-(\frac{\partial V}{\partial
T})_P(\frac{\partial P}{\partial V})_T.
\end{eqnarray}

That is why the statement that \((\frac{\partial S}{\partial
V})_T>0\) is therefore equivalent to the statement that the cell
expands when heated (at constant pressure). This behavior is
demonstrated by most solid bodies when heated. If we accept that
this is true (the cell expands when heated at constant pressure)
then we find  \(\delta V<0\), which was to be proved. We thus
conclude that the cell's size decrease. We consider our description
to be a general thermodynamic prerequisite that explains the
phenomenon of the Ling cell contraction, a prototype of muscle
contraction.

Before we consider the aggregation of key proteins in the Ling's
model, let us go back to the question of the transformation of the
internal energy of the cell when it dies. We have demonstrated above
that death occurs then
\begin{eqnarray}
(\delta E)_{T,V}<0.
\end{eqnarray}
If we use the symbol  \(\delta Q\) to denote the amount of heat that
entered the cell, we will find:
\begin{eqnarray}
\delta Q=(\delta E)_{T,V}+T(\frac{\partial S}{\partial V})_T\delta
V.
\end{eqnarray}
But we already know that \(\delta V\) and  have opposite signs, i.e.
\begin{eqnarray}
\delta Q<0.
\end{eqnarray}
In other words, the death and excitation of the cell are exothermic
reactions.

Let us consider now the thermodynamic prerequisites of the
aggregation of proteins that are present in a resting Ling's cell in
an expanded state.

Our analysis draws on the fact that the energy of an unfolded
protein molecule is greater than the energy of a folded protein
molecule and is based on the protein model in which the
configuration of the protein molecule can be characterized by a real
parameter \(x \in [0,1]\), describing the degree  to which the
molecule has unfolded. We assume that the energy of the protein
molecule is a function only of  \(x\)  and it is an increasing
function. We assume also that \(x = 0\) corresponds to a completely
folded molecule whose energy is at a minimum; and \(x = 1\)
corresponds to a completely unfolded molecule whose energy is at its
maximum.

Let us use the following supporting procedure. We will identify a
single protein molecule in the cell's protoplasm and call it
\(\mathfrak{X}\), and we'll use \(x\) as a parameter characterizing
the degree to which our single molecule is unfolded. If we call the
system's temperature \(T\), then all the thermodynamic equations
derived in the previous section can be applied to \(\mathfrak{X}\),
more particularly we have the equation:
\begin{eqnarray}
dE=TdS-hdx,\label{ttt}
\end{eqnarray}
for a certain function \(h(E, S)\) of the energy and entropy
\(\mathfrak{X}\). The value \(h(E, S)\) is a generalised force that
is thermodynamically conjugated to \(x\). When the external forces
that affect our protein molecule, are absent, \(h(E, S) = 0\). The
relation  (\ref{ttt}) is equivalent to the equation
\begin{eqnarray}
dE=TdS-PdV,
\end{eqnarray}
And that is why we can more or less directly apply to case the
conclusions  for the cell interacting with the environment, changing
\(x\mapsto V\), \(P\mapsto h\). So, Let's assume again that
\(s\mapsto s'=s-\delta s\), then
\begin{eqnarray}
(\delta x)_{T,h}=-T (\delta f)'(S)(\frac{\partial S}{\partial
h})_{T,x}.
\end{eqnarray}

We see that the sign of  \(\delta x\) is the same as the sign of
\((\frac{\partial S}{\partial h})_T\). But by virtue of the known
thermodynamic relation:
\begin{eqnarray}
(\frac{\partial S}{\partial h})_T=-(\frac{\partial x}{\partial
T})_h.
\end{eqnarray}
So, the sign \((\frac{\partial S}{\partial h})_T=-(\frac{\partial
x}{\partial T})_h\) will only be negative if the protein molecules
unfold and the cell temperature increases. So we have proved that
the protein molecules fold when the protoplasm is activated and when
the cell dies only if they unfold when the cell is heated. It is
clear that this statement is true only for the protein model used
here.

But the property of the protein molecule to unfold when heated
appears to be physically evident. This assumption seems to be
reasonable as the protein molecules are supposed to change to states
with more energy when heated, and the energy is an increasing
function of the parameter \(x\). That is why the protein molecule
folds under s \(s\mapsto s'=s-\delta s\), which was to be proved.

The question of how we apply this general method we used for the
analysis of more realistic protein models needs further research.

Let us consider now the question of why the Ling cell's
nonergodicity includes ion flows between the cell and the
environment when the physiological state of the Long cell is
changed. When analyzing this phenomenon we will restrict ourselves
to potassium ions that play an important role in the physiological
processes of the cell. We will try to answer the question of why
this positive ions are released by excited or dead cells. The answer
to this question will explain  another phenomenon well-known from
the classical cytology: why the damaged area in a cell always has
the negative charge to the adjoining intact areas that are not
affected by excitation or injury.

It is clear that our approach must be improved in order for us to be
able to consider a broader spectrum of ion characteristics, such as:
ion chemical nature, their physical properties, their concentration
in solution, etc.

 Let us assume that \(N\) means the number of
potassium ions in the cell \(\mathfrak{C}\). We can regard \(N\) as
a parameter, and it will be now clear why. Our analysis is carried
out in the thermodynamic limit and from this point of view the cell
can be considered very big. Let us attach a very small vessel
\(\mathfrak{S}\), to the cell which will be separated from it a
semipermeable membrane that is permeable only for potassium ions. We
will assume that in  \(\mathfrak{S}\) we have only potassium ions
and that  \(\mathfrak{S}\) is so small that the entropy of the ions
inside it is equal to zero. As a model for  \(\mathfrak{S}\) we can
take a very narrow potential well, the quantum energy levels of
which are not degenerate and the distance between adjoining energy
levels is much greater than the temperature \(T\). Then potassium
ions in \(\mathfrak{S}\) will be mostly in a basic energy state and
the thermodynamic probability \(W=1\), and \(S_{\mathfrak{S}}=\ln
W=0\). Let us also introduce an external electric field the
potential of which  \(\varphi\) everywhere except \(\mathfrak{S}\)
is equal to zero, and inside \(\mathfrak{S}\) is constant and equal
to \(\varphi_0\). Adiabatically changing \(\varphi_0\), we can pump
the potassium ions from \(\mathfrak{S}\) to the cell and back.
Meanwhile the entropy of the whole system will be constant and equal
to the entropy of the cell  \(S\). By virtue of the fact that
\(\varphi_0\) is a parameter in the Hamiltonian of the whole system,
\(((\delta S)_E)_{qs}=\rm const \mit\) during adiabatic change
\(\varphi_0\). Here \(E\) means the energy of the entire system. Let
us assume that \(E_{\mathfrak{C}}\) means the energy of the cell.
All the motion integrals that are related to the entire system
\(\mathfrak{S}+\mathfrak{C}\) are equally related to the cell
\(\mathfrak{C}\) because the potassium ions in  \(\mathfrak{S}\) are
immovable, and their coordinates and momentums can be removed from
the list of the dynamic variables. But
\begin{eqnarray}
(\delta S)_E=(\delta S)_{E_\mathfrak{C},N}+\frac{\partial
S}{\partial E_{\mathfrak{C}}}\delta E_\mathfrak{C}+\frac{\partial
S}{\partial N}\delta N.
\end{eqnarray}
But \(\delta E_\mathfrak{C}=-\varphi_0\delta N\), as the full energy
of the cell and of the vessel is constant. Considering that in
equilibrium:
\begin{eqnarray}
-\frac{\partial S}{\partial E_{\mathfrak{C}}}\varphi_0\delta
N+\frac{\partial S}{\partial N}\delta N=0,
\end{eqnarray}
we find:
\begin{eqnarray}
(\delta S)_E=(\delta S)_{E_\mathfrak{C},N}.
\end{eqnarray}
That is why during an adiabatic change in the parameter
\(\varphi_0\) or,  of the number of particles in the cell \(N\)
\begin{eqnarray}
((\delta S)_{E_\mathfrak{C},N})_{qs}=\rm const \mit,
\end{eqnarray}
Which means that \(N\) can be considered as a parameter.

Let us proceed now to the question of the amount of potassium ions
released by the cell during its activation and death. We assume that
the cell is immersed in a potassium ion solution and the chemical
potential of the potassium ions is constant and equal to \(\mu\).
The following thermodynamic relation will hold true:
\begin{eqnarray}
dE=TdS+\mu dN
\end{eqnarray}
and all its corollaries. This equation similar to the thermodynamic
equation
\begin{eqnarray}
dE=TdS-PdV,
\end{eqnarray}
if we equate  \(N\leftrightarrow V\), \(\mu\leftrightarrow -P\).
Therefore, to prove that the potassium ion yield from the cell takes
place when  \(s\mapsto s'=s-\delta s\) all we have to show is that
\begin{eqnarray}
(\frac{\partial S}{\partial N})_T>0.
\end{eqnarray}
To find \((\frac{\partial S}{\partial N})_T\) we will note that the
concentration of potassium ions in the cell is low and that is why
we can use the strong electrolyte theory (see [13]). This theory
suggests that in a low concentration of potassium ions the cell's
entropy (at fixed temperature) \(S=S_0+S^{K^+}+N\varphi(P,T)\),
where \(S_0\) - the cell entropy at zero ion concentration,
\(S^{K^+}\) - the entropy of only the potassium ions, calculated for
a perfect gas, and \(\varphi(P,T)\) is a function of the cell
temperature and pressure. That is why for the entropy of the
potassium ions we can use the standard formula (see [13]):
\begin{eqnarray}
S^{K^+}=N \ln \frac{e V}{N}-Nf'(T), \label{entropy}
\end{eqnarray}
where
\begin{eqnarray}
f(T)=-T \ln ((\frac{mT}{2 \pi})^{3/2}).
\end{eqnarray}
Here \(m\) is the mass of a potassium ion. It is clear that if the
potassium ion solution in the cell is thin enough \((\frac{\partial
S}{\partial N})_T>0\).

So, \(\delta N<0\), which was to be proved.

For simplicity of the analysis we assumed above that there are only
potassium ions in the cell. But in reality there are different kinds
of ions and all the derived formulas should be true for each kind of
ions. Particularly, during the activation and death of the cell all
kinds of ions should come out of the cell if their concentration is
low enough. But it is known that some ions, for example sodium ions,
are on the contrary, absorbed by the cell during activation or death
(according to Ling's model). . However, the cell becomes negatively
charged during activation, i.e. our explanation leads to the correct
answer "on average". This gives us hope that our explanation is
genuinely true and can be improved in a proper way.

To conclude of this section let us note that we considered the
resting state of the cell to be stationary non-equilibrium and the
state corresponding to the excitation and death to be equilibrium.
We considered these states as given and never asked the question why
there are transitions between these two states. One of the possible
hypotheses is that such transitions appear due to changes in the
conditions on the cell's boundaries. That is why in section 3 we
discussed in detail the role that the conditions on the system's
boundaries play in the establishment of thermodynamic equilibrium
inside the cell. The question of how true this hypothesis is will be
the subject of further research.

 \section{Conclusion.}
We used the results of [7] to analyse the Ling cell in this paper,
assuming that this kind of cell is a nonergodic system.

Using the property of nonergodicity as a starting point for our
research, we first discussed the nonergodicity of statistical
mechanics systems, we then developed the thermodynamics of
stationary non-equilibrium states assuming the existence of several
nontrivial first integrals that commutate between them. These
thermodynamics are applicable to any system in statistical
mechanics, including the Ling cell.

The analysis of the Ling cell properly demonstrated that using the
approach based on our thermodynamics of stationary non-equilibrium
we managed to derive such realistic properties of the excited Ling's
cell as heat generation, decrease of the cell size, the folding of
the proteins that are unfolded when the cell is resting and the
releasing of potassium ions. The applicability of the proposed
approach to the living cell model, such as the Ling cell, testifies
to the applicability and adequacy of our analysis for studying the
living state of matter.

We are very grateful to P. Agutter, A.V. Koshelkin, Yu. E. Lozovik,
A.V. Zayakin and E.N. Telshevskiy for valuable critical comments on
this article and very useful discussions.

\section{Appendix 1. Discussion of the inequality (\ref{TDuneq}).}

In this section we will try to justify the inequality
(\ref{TDuneq}).

Let us see the Hamiltonian dynamical system with \(n+k\) degrees of
freedom (\(k\ll n\)) and \(k\) independent first integrals in
involution \(K_1,...,K_k\). Locally we can always choose the
canonical coordinates \((p_1,q_1)\),...,\((p_{n+k},q_{n+k})\) in
such a way that \(K_1,...,K_n\) will only be the functions of
\((p_{n+1},q_{n+1})\),...,\((p_{n+k},q_{n+k})\). First, let us
assume that the coordinates that measure up this property can be
chosen globally. So, the phase space \(M\) of our system represents
as a direct product of the phase spaces   \(M=M_1\times M_2\), where
\(M_1\) corresponds to the coordinates
\((p_1,q_1)\),...,\((p_{n},q_{n})\), and \(M_2\) --- to the
coordinates \((p_{n+1},q_{n+1})\),...,\((p_{n+k},q_{n+k})\). Let us
denote through \(d \Gamma\) the element of the phase volume on
\(M\), and through \(d \Gamma^1\) and \(d \Gamma^2\) the elements of
the phase volumes on \(M_1\) and \(M_2\), correspondingly. Let us
assume that \(H(x,y)\) --- the Hamiltonian of our system, \(x \in
M_1\), \(y \in M_2\).

Let us assume that \(S(E)\) is the entropy calculated for the
microcanonical distribution and \(S_1(E,K_1',...,K_k')\) is the
entropy calculated for the generalized microcanonical distribution
corresponding to the \(K'_1,...,K'_k\) of the first integrals
\(K_1,...,K_k\). Let us recall that
\begin{eqnarray}
S_1(E,K_1',...,K_k')=\ln \int d\Gamma \delta(H-E)\prod
\limits_{i=1}^k\delta(K_i-K'_i).
\end{eqnarray}
We assume that \(S_1(E,K_1',...,K_n')\) measures up the equivalence
principle  (see section 4).

Locally on \(M_2\) we can build the function
\(\varphi_1,...,\varphi_k\), and do it in such a way that the set of
functions \((K_1,\varphi_1)\),...,\((K_k,\varphi_k)\) will represent
the set of canonical coordinates, i.e. the Poisson bracket in these
coordinates will have the standard form. In other words, we can
build a partition \(M_2\) in the area \(V_i\), \(i=1,2,...\) with a
piecewise-smooth boundary and in each of them we can choose the
functions \(\varphi_1^i,...,\varphi_k^i\), that satisfy the
requirement mentioned above. Let us assume that \(\Sigma\) --- a
certain joint surface of the level of the integrals \(K_1,...,K_k\),
and \(x \in \Sigma\). If (without losing generality) we assume that
\(\Sigma\) is connected, then the Hamiltonian phase flows generated
by \(K_1,...,K_k\) transitively acts on \(\Sigma\). And as these
phase flows keep the phase volume on \(M\) and the Hamiltonian, then
\begin{eqnarray}
S_1(E,K'_1,...,K'_k)=\ln \int d\Gamma_x^1 \delta(H(x,y)-E)+c,
\label{Rel}
\end{eqnarray}
where \(x \in M_1\),and \(c\) is the constant equal to:
\begin{eqnarray}
c=\ln \sum \limits_{i=1}^{\infty} \int \limits_{\Sigma\cap
V_i}d\varphi_1^i...d\varphi_k^i. \label{CC}
\end{eqnarray}
The constant \(c\) does not depend on \(E\) and can be discarded.

Let us now turn to the calculation of \((\delta S)_T\). If the
system \(M=M_1\times M_2\) is described by the Gibbs distribution
then the probability to find the system \(M\) at the point
\((x,y),\;x\in M_1,\;y\in M_2\) takes the form:
\begin{eqnarray}
w_{12}(x,y)=\frac{1}{Z(T)}e^{-\frac{H(x,y)}{T}}, \label{Cond}
\end{eqnarray}
where
\begin{eqnarray}
Z(T):=\int d \Gamma_x^1 d\Gamma_y^2 e^{-\frac{H(x,y)}{T}}.
\end{eqnarray}
The probability to find the system \(M_2\) at the point \(y\) is
given then by the expression:
\begin{eqnarray}
w_2(y)=\int d \Gamma_x^1 w_{12}(x,y).
\end{eqnarray}

Let us introduce the conditional probability \(w_{1|2}(x|y)\) to
find a system \(M_1\) at the point \(x\) at condition that the
system \(M_2\) is at the point \(y\) by the formula:
\begin{eqnarray}
w_{12}(x,y)=w_{1|2}(x|y)w_2(y). \label{Cond1}
\end{eqnarray}

Von Neumann entropy of the system \(M\) takes the form:
\begin{eqnarray}
S(E(T))=-\int d \Gamma_x^1 d \Gamma_y^2 w_{12}(x,y) \ln w_{12}(x,y).
\end{eqnarray}

Using (\ref{Cond1}) \(S(E(T))\) we can transform to the form:
\begin{eqnarray}
S(E(T))=\langle S_1(E(T),y)\rangle_{M_2}+S_2(T),
\end{eqnarray}
where
\begin{eqnarray}
S_1(E(T),y)=-\int d \Gamma_x^1 w_{1|2}(x|y) \ln w_{1|2}(x|y),
\end{eqnarray}
\(\langle\cdot\rangle_{M_2}\) means the averaging over \(M_2\) by
the measure \(w_2(y) d \Gamma_y\), and
\begin{eqnarray}
S_2(T)=-\int d \Gamma_y^2 w_2(y) \ln w_2(y).
\end{eqnarray}
Note that only averaged values can be measured at experiment. We
find:
\begin{eqnarray}
(\delta S)_T=S_2(T).
\end{eqnarray}
At that we meant the formula (\ref{Rel}) and the fact that within
the thermodynamic limit the descriptions by the microcanonical and
canonical assemblies coincide. At the same time we also consider, of
course, that the constant \(c\) from the formula (\ref{CC}) does not
depend on \(K'_1\),...,\(K'_k\). When we will proceed to the
discussion of the general case that will not require the
presentability of \(M\) in the form of the direct product \(M_1\)
and \(M_2\) we will see that this assumption is justified.

Let us try to describe the dynamics of the system \(M_2\), using the
fact that the number of degrees of freedom \(M_2\) \(n\) is much
greater that the number of degrees of freedom \(M_1\) \(k\). The
Hamiltonian canonical equations for \(M_2\) take the form:
\begin{eqnarray}
\dot{p}_i=-\frac{\partial H(p_1,q_1,...,p_{n+k},q_{n+k})}{\partial
q_i},\nonumber\\
\dot{q}_i=\frac{\partial H(p_1,q_1,...,p_{n+k},q_{n+k})}{\partial
p_i},\nonumber\\
i=n+1,...,n+k.
\end{eqnarray}
As the number of degrees of freedom \(M_2\) is much less that the
number of degrees of freedom \(M_1\) we can average the right parts
of the last equation by \(w_{1|2}(x,y)\). As a result, skipping some
quite trivial calculations we will find:
\begin{eqnarray}
\dot{p}_i=-\frac{\partial
F(p_{n+1},q_{n+1},...,p_{n+k},q_{n+k}|T)}{\partial
q_i},\nonumber\\
\dot{q}_i=\frac{\partial
F(p_{n+1},q_{n+1},...,p_{n+k},q_{n+k}|T)}{\partial
p_i},\nonumber\\
i=n+1,...,n+k,
\end{eqnarray}
where

\begin{eqnarray}
F(p_{n+1},q_{n+1},...,p_{n+k},q_{n+k}|T)=-T\ln
Z_1(y|T),\nonumber\\
Z_1(y|T)=\int d \Gamma_x^1 e^{-\frac{H(x,y)}{T}}, \nonumber\\
y=(p_{n+1},q_{n+1},...,p_{n+k},q_{n+k}).
\end{eqnarray}

So, this way the dynamics of the system \(M_2\) will be described by
the canonical Hamiltonian equations with the Hamiltonian \(F(y|T)\).

Let us note that this way the defined Hamiltonian is not defined
uniquely but accurate to the temperature arbitrary function
\(f(T)\). Let us define the new Hamiltonian of the system  \(M_2\)
\begin{eqnarray}
H_2(y|T)=F(y|T)+f(T),
\end{eqnarray}
in such a way that makes true the following:
\begin{eqnarray}
(\delta E)_T=\langle H_2(y|T)\rangle_{M_2}.
\end{eqnarray}
Let us show that in this case
\begin{eqnarray}
\langle\frac{d}{dT} H_2(y|T)\rangle_{M_2}=0. \label{aaa}
\end{eqnarray}
For this we will proceed from the identical equation:
\begin{eqnarray}
\frac{\frac{d}{dT} (\delta S)_T}{\frac{d}{dT} (\delta
E)_T}=\frac{1}{T}.
\end{eqnarray}
But

\begin{eqnarray}
(\delta S)_T=S_2(E(T))
=-\int d \Gamma_y^2 w_2(y) \ln w_2(y).\nonumber\\
\frac{d}{dT}S_2(E(T))=\int d \Gamma_y^2 \frac{H_2(y|T)}{T}
\frac{d}{dT} \frac{e^{-\frac{H_2(y|T)}{T}}}{Z_2(T)},
\end{eqnarray}
where
\begin{eqnarray}
Z_2(T):=\int d \Gamma_y^2 e^{-\frac{H_2(y|T)}{T}}.
\end{eqnarray}
We have:
\begin{eqnarray}
\frac{d}{dT}S_2(E(T))=-\langle\Delta(\frac{H_2(y|T)}{T})\Delta(\frac{d}{dT}\frac{H_2(y|T)}{T})\rangle_{M_2},
\label{174}
\end{eqnarray}
where we assumed by definition:
\begin{eqnarray}
\Delta g(y):=g(y)-\langle f(y) \rangle_{M_2}.
\end{eqnarray}
Making the similar calculations, we find:
\begin{eqnarray}
\frac{d}{dT}\langle H_2(y|T)\rangle=\langle \frac{d}{dT}
H_2(y|T)\rangle-\langle\Delta({H_2(y|T)})\Delta(\frac{d}{dT}\frac{H_2(y|T)}{T})\rangle_{M_2}.
\label{176}
\end{eqnarray}
Comparing (\ref{174}) and (\ref{176}) we will find:
\begin{eqnarray}
\langle\frac{d}{dT} H_2(y|T)\rangle_{M_2}=0,
\end{eqnarray}
which was to be proved.

Let us describe now the effective Hamiltonian of the whole system
\(M_1\times M_2\). The dynamics of the system \(M_1\) can be
described by the Hamiltonian \(H(x,y)\) where \(y \in M_2\) and
depends on time, and \(x\in M_1\). But because of \(K_1,...,K_k\)
are the motion integrals, the value \(H(x,y)\) at fixed \(x\) is the
same for all \(y\) corresponding to the same value of the motion
integrals. That is why the system \(M_1\) is described by the
Hamiltonian \(H(x,y)\), where \(y\) is constant with time and is
chosen the same way as in the formula (\ref{Rel}).

Further the effective Hamiltonian of the system \(M_2\) \(H_2(y|T)\)
depends on the temperature \(T\) of the system \(M_1\), but
temperature is an intensive parameter and instead of \(T\) we can
use the energy of the system \(M_1\), falling within one degree of
freedom. Let us assume by definition:
\begin{eqnarray}
H'_2(y|\frac{E_1}{n})=H_2(y|T),
\end{eqnarray}
where \(E_1\) --- the energy of the system \(M_1\).

As an effective Hamiltonian of the system \(M=M_1\times M_2\) the
following expression will be used:
\begin{eqnarray}
H_{eff}(y,z)=H(y,x)+H'_2(z|\frac{H(y,x)}{n}). \label{eff}
\end{eqnarray}
Let us recall that according to this formula \(x\) is chosen the
same way as in (\ref{Rel}).

Let us show that within the limit \(n\rightarrow \infty\) the
Hamiltonian equation for \(H_{eff}\) will coincide with the
Hamiltonian equations for the initial Hamiltonian if the point \(x\)
is on the same joint surface of the level of the integrals
\(K_1,...,K_k\), that \(z\).

Let us assume that \((y,z) \in M=M_1\times M_2\). Let us analyze the
derivative with time of the point \(z\) by virtue of the Hamiltonian
equations for the Hamiltonian \(H_{eff}\). This will be exactly the
derivative by virtue of the Hamiltonian motion equations
corresponding to the Hamiltonian \(H_2(z|T)\) and the fact that this
derivative coincide with the derivative with time by virtue of the
Hamiltonian motion equations built by \(H(y,z)\) was shown by us
above. Let us assume now that \(R(y)\) is the dynamic variable on
\(M_1\). Let us calculate its derivative with time by virtue of
\(H_{eff}\). We have:
\begin{eqnarray}
\dot{R}(y)=(H(y,x),R(y))(1+\frac{1}{n}H^{''}_2(z,\frac{H(y,x)}{n})).
\label{123}
\end{eqnarray}
Here the symbol \(H^{''}_2(\varepsilon,z)\) denotes the derivative
\(H^{'}_2(\varepsilon,z)\) by the first argument. Within the limit
\(n \rightarrow \infty\) the right part of (\ref{123}) evidently
transforms into \((H(y,x),R(y))\), i.e. within this limit
\begin{eqnarray}
\dot{R}(y)=(H(y,x),R(y)),
\end{eqnarray}
which was to be proved.

Let us note as well that \(H_{eff}\) depends on the choice of the
point \(x \in M_2\). However, because of the equivalence principle,
the entropic properties of the system \(M=M_1\times M_2\) will not
depend on this choice and these last ones are the only ones that are
important for us. That is why from the point of view of the
calculations of the entropic properties the Hamiltonian
\(H_{eff}(y,z)\) is as good as the initial \(H(y,z)\) and we will
further use only the first one.

Let us note finally that the effective Hamiltonian with the form of
(\ref{eff}) is defined accurate to the arbitrary function
\(f(\frac{H(y,x)}{n})\). I.e. the effective Hamiltonians
corresponding to the different choices of such normalization induce
the same motion equations. That is why all the thermodynamic
characteristics calculated by \(H_{eff}\) corresponding different
choices of the normalization must coincide. And this is really so.
Actually at the old temperature the adding of
\(f(\frac{H(y,x)}{n})\) to \(H_{eff}\) the arbitrary function  can
be considered by the corresponding change \(H_1(y,x)\). But at the
same time with \(n \rightarrow \infty\) the average value of the
energy \(E_1\) of the system \(M_1\) will change by the value \(\sim
1\), and that is why the entropy of the system \(M_2\) will change
by the value \(\sim \frac{1}{n}\), i.e. it will not change at all
within the thermodynamic limit.

Let us use \(S(T)\) to denote the entropy of the system \(M\), as
the function of the temperature \(T\), calculated using Gibbs
distribution for the Hamiltonian \(H_{eff}\). Let us use \(E_1\) to
denote the average \(H_1(y,x)\), and \(E_2\) to denote the average
\(H_2(y|T)\) by the measures on \(M_1\) and \(M_2\) respectively,
induced by the same Gibbs distribution. Let us assume also
\(E=E_1+E_2\). The direct calculation shows that \(\frac{d
S}{dE}=\frac{1}{T}\). From the other point of view we normalized
\(H_{2}(y|T)\) in such a way that \(\frac{d S_2(T(E_2))}{d
E_2}=\frac{1}{T}\) would be true. From these two equalities we
obtain \(\frac{d S_1(E_1)}{d E_1}=\frac{1}{T}\). I.e. in the
thermodynamic equilibrium state the temperatures of the systems
\(M_1\) and \(M_2\) coincide, and this equality is true precisely
and not only in the thermodynamic limit \(k=\rm const \mit\), \(n
\rightarrow \infty\).

If \(k\) is very large (being at the same time much less than
\(n\)), then according to the main principles of the statistical
mechanics, instead of the description of \(M_2\) by means of Gibbs
canonical distribution we can use the description by means of the
microcanonical distribution and the function of the distribution for
the whole system \(M\), at which the energies of the systems \(M_1\)
and \(M_2\) will be equal to \(E_1\) and \(E_2\) respectively, will
take the form:
\begin{eqnarray}
\delta(H(y,x)-E_1)\delta(H_2'(z|\frac{E_1}{n})-E_2),
\end{eqnarray}
\(y \in M_1,\; z\in M_2\). The entropies of the systems \(M_1\) and
\(M_2\) will be equal to
\begin{eqnarray}
S_1(E_1)=\ln \int \delta(H(y,x)-E_1)  d \Gamma_y^1,\nonumber\\
S_2(E_2,E_1)=\ln  \int \delta(H_2'(z|\frac{E_1}{n})-E_2) d
\Gamma_z^2.
\end{eqnarray}

To obtain the inequality that we need let us make the following
conceptual experiment. Let us make the system \(M_2\) contact the
thermostat \(T\), which is at the temperature \(T\). As for the
system \(M_1\), we will take a very large system \(M_3\) that will
be connected to the system \(M_1\) via a heat pump \(N\), working by
Carnot cycle. We will consider the working body of this pump to be
so small that its thermal capacity can be neglected. Let us also
assume that during the thermal contact with \(M_1\) or \(M_3\) the
temperature of the working body of the pump is equal to the
temperature of \(M_1\) or \(M_3\) (depending on which one has
thermal contact with the working body of the pump), i.e. the heat
pumping is done without the entropy increase. The working body of
the pump is connected to a shaft by something like a connecting rod
gear, this shaft can rotate and on the axis of which there is a
spring \(P\), possessing the energy \(E_P\).

Let us note that because of our assumptions about the effective
Hamiltonian  \(H_{eff}(y,z)\) the dynamics given by it on \(M_1\)
are exactly the dynamics given by \(H(y,x)\) and that is why the
system \(M_1\) can be considered closed. Or more precisely as the
system \(M_1\) is related to \(M_3\) by means of the heat pump \(N\)
the system that consists from \(M_1\), \(M_3\), \(N\), \(P\) can be
considered closed. We will now assume that the elastic
characteristics of the spring \(P\) are picked the way that the work
that must be done to the working body \(N\) is exactly equal to the
change of the spring energy. So this way we have a whole continuum
of equilibrium states of the system consisting of \(M_1\), \(M_3\),
\(N\), \(P\), which is parametrized by, for example, the angle of
rotation of the pump shaft and at \(S_1+S_3=\rm const \mit\),
\(E_1+E_3+E_P=\rm const \mit\).

When our whole system obtains the equilibrium then the temperature
of the thermostat is equal to the temperature of \(M_1\) and the
temperature of \(M_3\). From the condition of maximal entropy of the
whole system that consists of \(M_1\), \(M_2\), \(M_3\), \(T\),
\(N\), \(P\) at fixed full energy considering the equalities:
\(S_1+S_3=\rm const \mit\), \(E_1+E_3+E_P=\rm const \mit\) we will
find that in the equilibrium the following value
\begin{eqnarray}
S_2(E_2,E_1)+S_T(E_T)
\end{eqnarray}
should reach the maximum at the additional condition
\begin{eqnarray}
E_2+E_T=\rm const \mit.
\end{eqnarray}
Here \(E_T\) --- the energy of the thermostat and \(S_T(E_T)\) ---
the entropy of the thermostat.

Instead of the dependence \(S_2(E_2,E_1)\) on \(E_1\) it appears
more convenient to consider the dependence \(S_2\) from the system
temperature \(M_1\) \(\lambda(E_1)\). Let us introduce a new
function
\begin{eqnarray}
S'(E,\lambda(E_1)):=S_2(E,E_1).
\end{eqnarray}

The condition of the maximality of the entropy of the system \(M_2\)
and the thermostat lead to the equalities:
\begin{eqnarray}
\frac{\partial
S'(E,\lambda)}{\partial\lambda}|_{\lambda=T,\;E=E(T)}=0,\nonumber\\
\frac{\partial S'(E,\lambda)}{\partial
E}|_{\lambda=T,\;E=E(T)}=\frac{1}{T}. \label{Nor}
\end{eqnarray}
But these equalities can be obtained directly. The second equality
is obtain from the ordinary formulas of the statistical mechanics if
we assume that the system \(M_2\) is described by the Gibbs
canonical distribution with Hamiltonian \(H_2(y,\lambda)\). Let us
set up the first equality. Let us adiabatically change the parameter
\(\lambda\), \(\lambda\mapsto T+\delta\lambda\). At such change of
the parameter \(\lambda\) the entropy of the system \(M_2\) will not
change. But by virtue of (\ref{aaa}) the energy \(M_2\) will not
change. I.e. the derivative of the entropy of the system \(M_2\) by
\(\lambda\) at fixed energy is equal to zero, which is exactly the
equality that we needed.

But for the maximality of the entropy of the system \(M_2\) and the
thermostat it is also necessary that the matrix of the second
derivatives \(S'(E,\lambda)\) at the point \(E=E(T),\lambda=T\)
would be defined negative.

To prove now the inequality (\ref{TDuneq}), we should prove that
\begin{eqnarray}
\frac{d^2}{dE^2}S'(E,T(E))\leq 0.
\end{eqnarray}
But,
\begin{eqnarray}
\frac{d^2}{dE^2}S'(E,T(E))=\frac{\partial^2 S'(E,T)}{\partial
E^2}+2\frac{d T(E)}{dE}\frac{\partial^2 S'(E,T)}{\partial E\partial
T}+\nonumber\\
+(\frac{d T(E)}{dE})^2\frac{\partial^2 S'(E,T)}{\partial ^2 T}\leq
0,
\end{eqnarray}
And the last inequality is true by virtue of the negative
determination of the matrix of the second derivatives
\(S'(E,\lambda)\) at the point \(E=E(T),\lambda=T\). Which was to be
proved.

In a general case when \(M\) is impossible to be conceived as a
direct product \(M=M_1\times M_2\) it can be shown that for some
covering \((\tilde{M},\tilde{H})\) of our Hamiltonian system
\((M,H)\) such factorization is possible. Deriving from the very
beginning all the thermodynamic relations for
\((\tilde{M},\tilde{H})\), as it was done in section 3, we come
again to the inequality (\ref{TDuneq}).

Or more exactly we want to say the following. In all our
thermodynamic analysis the Hamiltonian \(H\) and the integrals
\(K_1,...,K_k\) depended on a certain parameter \(\lambda\) (
volume), \(H(\lambda)\), \(K_1(\lambda),...,K_k(\lambda)\). Further
the dynamic variable \(X(\lambda)\), depending on the parameter
\(\lambda\) at \(\lambda=0\) will be denoted simply as \(X\).

We can build a covering Hamiltonian system \((\tilde{M},\tilde{H})\)
of the system \((M,H)\), so that \(\tilde{M}\) would be conceived as
a direct product \(\tilde{M}=\tilde{M}_1\times\tilde{M}_2\) and so
that will measure up one more additional condition. Let us assume
that \(\pi\) is a canonical projection of \(\tilde{M}\) on \(M\) and
let us assume that \(\tilde{H}(\lambda)\),
\(\tilde{K}_1(\lambda)\),...,\(\tilde{K}_k(\lambda)\) are the lift
\(H(\lambda)\), \(K_1(\lambda)\),...,\(K_k(\lambda)\) on
\(\tilde{M}\). The additional condition mentioned above is that the
canonical coordinates on \(\tilde{M}_2\) are
\(\tilde{K}_1,...,\tilde{K}_k\) and the conjugated variables
\(\varphi_1,...,\varphi_k\), and \(\forall i=1,...,k\) \(\varphi_i\)
runs all the real axis.

We will now derive all the thermodynamic relations for
\(\tilde{M},\tilde{H}\). But here we get the difficulty related to
the fact that \(\tilde{M}_2\) is not compact. There is how we
propose to overcome this difficulty. We propose to deal with
unnormalized distributions \(w\) for which the following is true
\(\int \limits_{\tilde{M}}w(x)d \tilde{\Gamma}_x=\infty\), where \(d
\tilde{\Gamma}_x\) is the element of the phase volume on
\(\tilde{M}\).

Let us assume that \(R_L=\{x \in \tilde{M}|\forall i=1,...,k\;
|\varphi_i(x)|<L\}\). To calculate the average of the dynamic
variable \(\tilde{f}\) on \(\tilde{M}\) by \(w\) we propose to use
the following formula:
\begin{eqnarray}
\langle\tilde{f}\rangle=\lim \limits_{L\rightarrow \infty}
\frac{\int \limits_{R_L} \tilde{f}(x)w(x)d\tilde{\Gamma}_x}{\int
\limits_{R_L} w(x)d\tilde{\Gamma}_x}. \label{Lim}
\end{eqnarray}

The existence of the limit (\ref{Lim}) --- a nontrivial question but
we are going to show that in all the cases we are interested the
limit (\ref{Lim}) exists.

We will say that the function \(\tilde{f}\) on \(\tilde{M}\) is
periodical if \(\forall x,y \in \tilde{M}\) such that
\(\pi(x)=\pi(y)\) \(\tilde{f}(x)=\tilde{f}(y)\). All the lifts of
the functions given on \(M\) on \(\tilde{M}\) are periodical and all
the periodical functions can be obtained this way. All the
interesting for us dynamic variables on \(\tilde{M}\) appear exactly
as lifts of the functions given on \(M\) and that is why are
periodical. Further the generalized microcanonical distributions on
\(\tilde{M}\), built by
\(\tilde{H}(\lambda),\;\tilde{K}_1(\lambda),...,\tilde{K}_k(\lambda)\)
are also periodical. We are going to demonstrate that if
\(\tilde{f}(x)\) and \(w(x)\) are periodical functions then the
limit(\ref{Lim}) exists.

Let us assume that \(d\tilde{\Gamma}^1\) is the element of the phase
volume on \(\tilde{M}_1\). Let us assume that
\(\tilde{M}_{K'_1,...,K'_k}\) is the surface of the level on
\(\tilde{M}\), corresponding to the values \(K'_1,...,K'_k\) of the
integrals \(\tilde{K}_1,...,\tilde{K}_k\). Let us assume that
\(d\tilde{\nu}=d\tilde{\Gamma}^1 d\varphi_1,...,d\varphi_k\) is the
measure on \(\tilde{M}_{K'_1,...,K'_k}\). The projection\(\pi\)
reflects \(\tilde{M}_{K'_1,...,K'_k}\) on \({M}_{K'_1,...,K'_k}\) is
the surface of the level on \(M\) corresponding to the values
\(K'_1,...,K'_k\) of the integrals \(K_1,...,K_k\). On
\({M}_{K'_1,...,K'_k}\) we have a measure \(d\nu\), That can be
characterized as a quotient of the phase volume \(d \Gamma\) by
\(dK_1...dK_k\): \(d\Gamma=d\nu dK_1...dK_k\). It is evident that
the measure \(d\tilde{\nu}\) is a lift of the measure \(d{\nu}\)
from \({M}_{K'_1,...,K'_k}\) to \(\tilde{M}_{K'_1,...,K'_k}\) by
virtue of the canonical projection \(\pi\). To prove the conclusion
about the existence of the limit (\ref{Lim}), it is evidently enough
to prove that for any \(K'_1,...,K'_k\) the following limit
\begin{eqnarray}
\lim \limits_{L \rightarrow \infty} \frac{1}{L^k} \int
\limits_{R_L\cap \tilde{M}_{K'_1,...,K'_k}} \tilde{f}(x) d
\tilde{\nu}(x)
\end{eqnarray}
exists for any (good enough) periodical function
\(\tilde{f}(x)=f\circ\pi(x)\). And for this it is enough, evidently,
to prove that for any differentiable function \(\psi\) on
\(\tilde{M}_1\) with a compact support there is a limit:
\begin{eqnarray}
\lim \limits_{L \rightarrow \infty} \frac{1}{L^k} \int
\limits_{\tilde{M}_1}\psi(x)d\Gamma^1_x \int
\limits_{|\varphi_i|<L}\tilde{f}(x,\varphi_1,...,\varphi_k).
\end{eqnarray}
But on second thought, the existence of the last limit succeeds
directly from \(k\)- dimensional analogue of the von Neumann ergodic
theorem [14], applied to the commutated flows on
\({M}_{K'_1,...,K'_k}\), that are induced by the Hamiltonians
\(K_1,...,K_k\) keeping the measure \(d\nu\) and to restriction of
the function \(f\) on \({M}_{K'_1,...,K'_k}\).

Now in order to establish all the necessary to us thermodynamic
relations, for example the relation \(dE=TdS-PdV\), we derive them
for \(\tilde{M}\) cut of \(R_L\), and then tend \(L\) to infinity.
The only difficulty that can occur is the divergence of the entropy
at \(L\rightarrow \infty\), but this divergence is removed by the
diminution of the logarithmically divergent constant\(k \ln L\) from
the entropy and this constant is the same at any value of the
parameter \(\lambda\).

The inequality (\ref{TDuneq}) that we need is now derived as before
but at \(\lambda=0\). It is enough to obtain again all our
conclusions concerning the Ling cell.

\section{Appendix 2. Discussion of the boundary condition \ref{Gran}.}
The aim of this appendix is to make more or less obvious the
existence of such temperature \(T_0\), above which \((\delta
S)_E=\rm const \mit\).

Let us assume that at \(s\mapsto s+\delta s\) the integral\(K\)
becomes inactive. The entropy of the system \(S(E,K')\) at fixed
energy \(E\) depends on the value \(K'\) of this integral and the
values of the other integrals that we will not name here. When
discussing the equivalency principle we said that \(S(E,K')\) should
not depend on \(K'\), but it would be more precise to say that
\(S(E,K')\) reaches the maximum on some interval \([\alpha,\beta]\)
of the values \(K\), and decreases drastically outside it. It is
clear that \((\delta S)_E =\ln (\beta-\alpha)\). But this kind of
dependence \(S(E,K')\) from \(K'\) is not analytical and is proper
to the systems with an infinite number of degrees of freedom, i.e.
in the thermodynamic limit.

As an example of non-analyticity within the thermodynamic limit we
can see the phase transition liquid \(\leftrightarrow\) vapor. Let
us see the dependence of the system pressure from its volume at
temperature lower that critical. When the volume is decreasing from
infinity the system pressure first begins to increase and then on
some interval of volume change will remain constant. This interval
corresponds to the co-existence of the liquid and vapor phases. At
further volume decrease the pressure will increase again. So this
way this isotherm is not describe by the analytical function. When
temperature is increasing the analyticity is restored. The
temperature at which the analyticity is restored is called critical.
Above this temperature the difference between vapor and liquid
disappears.

Another example of the non-analytical behavior of the thermodynamic
values which is more close to the Ling cell is borrowed from the van
der Waals theory of the second-order phase transitions [13]. The
second-order phase transition is usually related to some symmetry
breaking and the breaking of this symmetry is described by a certain
parameter of the order \(\eta\). We will analyze here the simplest
case of the real-valued \(\eta\) and \(\mathbb{Z}_2\)-symmetry that
only changes the sign \(\eta\); \(\eta\mapsto-\eta\).

In the neighborhood of the point of phase transition the Taylor
approximation of the thermodynamic potential \(\Phi(\eta,T)\)
accurate to the inappreciable terms with higher orders, takes the
form:
\begin{eqnarray}
\Phi(\eta,T)=c_0\eta^4+c_1(T-T_c)\eta^2,
\end{eqnarray}
where \(T_c\) --- critical temperature and \(c_0\), \(c_1\) ---
certain positive constants.

At \(T\geq T_c\) \(\Phi(\eta,T)\) has by \(T\) one minimum in zero.
At \(T<T_c\) \(\Phi(\eta,T)\) has one local maximum by \(\eta\) in
zero and two minimums in the points
\begin{eqnarray}
\eta_{1,2}=\pm \sqrt{\frac{c_1}{2c_0}(T_c-T)}. \label{1,2}
\end{eqnarray}
We assume that the parameter of the order \(\eta\) is additive and
that is why is lower that critical temperature, the analyzed system
is separated into domains and in each of them \(\eta\) takes one of
the two opposite values, corresponding to the minimum of the
thermodynamic potential of the domain. The resulting value \(\eta\)
will be bounded between two values \(\eta_1\) and \(\eta_2\). From
the other side, as the domains are big enough we can neglect the
interaction between the different domains when calculating the
thermodynamic potential and the thermodynamic potential will be
equal to the minimal value of \(\Phi(\eta,T)\) by \(\eta\).

So this way we come to conclusion that instead of \(\Phi(\eta,T)\)
we should use its upper envelope \(\tilde{\Phi}(\eta,T)\) (by
\(\eta\)) and at \(T>T_c\) \(\tilde{\Phi}(\eta,T)\) will have the
plateau, i.e. \(\tilde{\Phi}(\eta,T)\) will not be the analytic
function anymore.

The recovery of the analyticity at the temperature increase is
explained in the following way. At the temperatures high enough the
kinetic energy of every particle becomes much more than the
potential energy of the particle interaction and the particles can
be considered non-interacting. I.e. at high temperatures all the
particles that compose the system can be considered free and for
such system all the characteristics must be analytical.

The same way we will assume that at the temperature high enough
\(S(E,K')\) will be the analytical function of the parameter \(K'\).
Then it is either constant or reaches the maximum at an isolated
point. But \((\delta S)_E =\ln (\beta-\alpha)\), as it was shown
above, decreases with the temperature increase, i.e. at temperature
high enough \(S(E,K')\) can not be constant on the whole interval of
the change of \(K'\). I.e. \(S(E,K')\) reaches the maximum at an
isolated point and there is no more difference between using the
generalized microcanonical distribution or just the microcanonical
distribution.

\end{document}